# Learning and innovative elements of strategy adoption rules expand cooperative network topologies

Shijun Wang[1,*], Máté S. Szalay[2], Changshui Zhang[1], Peter Csermely[2, #]

**1** Department of Automation, FIT Building, Tsinghua University, 100084 Beijing, China, **2** Department of Medical Chemistry, Semmelweis University, Puskin str. 9, H-1088 Budapest, Hungary

**Cooperation plays a key role in the evolution of complex systems. However, the level of cooperation extensively varies with the topology of agent networks in the widely used models of repeated games. Here we show that cooperation remains rather stable by applying the reinforcement learning strategy adoption rule, Q-learning on a variety of random, regular, small-word, scale-free and modular network models in repeated, multi-agent Prisoner's Dilemma and Hawk-Dove games. Furthermore, we found that using the above model systems other long-term learning strategy adoption rules also promote cooperation, while introducing a low level of noise (as a model of innovation) to the strategy adoption rules makes the level of cooperation less dependent on the actual network topology. Our results demonstrate that long-term learning and random elements in the strategy adoption rules, when acting together, extend the range of network topologies enabling the development of cooperation at a wider range of costs and temptations. These results suggest that a balanced duo of learning and innovation may help to preserve cooperation during the re-organization of real-world networks, and may play a prominent role in the evolution of self-organizing, complex systems.**

## Introduction

Cooperation is necessary for the emergence of complex, hierarchical systems [1–5]. Why is cooperation maintained, when there is a conflict between self-interest and the common good? A set of answers emphasized agent similarity, in terms of kin- or group-selection and compact network communities, which is helped by learning of successful strategies [2,3]. On the other hand, agent diversity in terms of noise, variation of behavior and innovation, as well as the changing environment of the agent-community all promoted cooperation in different games and settings [3,6–8].

*Present address: Diagnostic Radiology Department, Clinical Center, National Institutes of Health, 10 Center Dr, MSC 1182, Bldg 10, Room B2S-231, Bethesda, MD 20892-1182 USA.

#To whom correspondence should be addressed. E-mail: csermely@puskin.sote.hu

This article contains electronic supplementary material.

Small-world, scale-free or modular network models, which all give a chance to develop the complexity of similar, yet diverse agent-neighborhoods, provide a good starting point for the modeling of the complexity of cooperative behavior in real-world networks [9–13]. However, the actual level of cooperation in various games, such as the Prisoner's Dilemma or Hawk-Dove games is very sensitive to the topology of the agent network model [14–16, Electronic supplementary material 1 – ESM1 – Table S1.1]. In our work we applied a set of widely used network models and examined the stability of cooperation after repeated games using the reinforcement learning strategy adoption rule, Q-learning. To examine the surprising stability of cooperation observed, when using Q-learning, we approximated the complex rules of Q-learning by designing a long-term versions of the best-takes-over and other strategy adoption rules as well as introducing a low level of randomness to these rules. We found that none of these features alone results in a similar stability of cooperation in various network models. However, when applied together, long-term ('learning') and random ('innovative') elements of strategy adoption rules can make cooperation relatively stable under various conditions in a large number of network models. Our results have a wide application in various complex systems of biology from the cellular level to social networks and ecosystems.

## Results

**Sensitivity of cooperation on network topology**
As an illustrative example for the sensitivity of cooperation on network topology, we show cooperating agents after the last round of a 'repeated canonical Prisoner's Dilemma game' (PD-game) on two, almost identical versions of a modified Watts-Strogatz-type small-world model network [13,17]. Comparison of the top panels of Figure 1 shows that a minor change of network topology (replacement of 37 links from 900 links total) completely changed both the level and topology of cooperating agents playing with a best-takes-over short term strategy adoption rule. We have observed a similar topological sensitivity of cooperation in all combinations of (a) other short-term strategy adoption rules; (b) a large number of other network topologies; (c) other games, such as the extended Prisoner's Dilemma or Hawk-Dove games (ESM1 Figures S1.1 and S1.6).

**Q-learning stabilizes cooperation in different network topologies**
On the contrary to the general sensitivity of cooperation to the topology of agent-networks in PD-games using the short-term strategy adoption rule shown above, when the long-term, reinforcement learning strategy adoption rule, Q-learning was applied, the level and configuration of cooperating agents showed a surprising stability (cf. the bottom panels of Figure 1). Just oppositely to the short-term strategy adoption rule shown on the top panels of Figure 1, the Q-learning strategy adoption rule (a) is based on the long-term experiences of the agents from all previous rounds allowing some agents to choose a cooperative strategy despite of the current adverse effects, and (b) is an 'innovative' strategy adoption rule [3] re-introducing cooperation even under conditions, when it has already been wiped out from the network-community completely [18,19].

Extending the observations shown on Figure 1 we decided to compare the level of cooperation in PD-games on small-world and scale-free networks at various levels of temptations (T, the defector's payoff, when it meets a cooperator) in detail. The top panel of Figure 2 shows that the cooperation level of agents using the best-takes-over strategy adoption rule rapidly decreased with a gradual increase of their temptation to defect. This was generally true for both small-world, and scale-free networks leaving a negligible amount of cooperation at T-values higher than 4.5. However, at smaller temptation levels the level of cooperation greatly differed in the two network topologies. Initially, the small-world network was preferred, while at temptation values higher than 3.7, agents of the scale-free network developed a larger cooperation. The behavior of agents using the Q-learning strategy adoption rule was remarkably different (top panel of Figure 2). Their cooperation level remained relatively stable even at extremely large temptation values. Moreover, the cooperation levels of agents using Q-learning had no significant difference, if we compared small-world and scale-free networks. This behavior continued at temptation values higher than 6 (data not shown). We have observed the same differences in both the extent of cooperation at extremely high temptations (or gains of hawks meeting a dove in the Hawk-Dove game) and the topological sensitivity of cooperation in all combinations of (a) other short-term strategy adoption rules; (b) a large number of other network topologies; (c) other games, such as the extended Prisoner's Dilemma or Hawk-Dove games (ESM1 Figures S1.2 and S1.6).

**Long-term strategy adoption rules improve but do not stabilize cooperation in different networks**

Next we wanted to see, if other long-term strategies besides Q-learning can also promote cooperation between agents. In Q-learning agents consider a long-term experience learned in all the past rounds of the play. Therefore, we modified the best-takes-over strategy adoption rule allowing the agents to use accumulative rewards of their neighbors in all past rounds instead of the reward received just in the last round. In agreement with our expectations, both on small-world and scale-free networks this long-term strategy adoption rule outperformed its short-term version allowing a larger number of agents to cooperate – especially at high temptation values. Importantly, the differences between cooperation levels observed in small-world and scale-free networks were even greater, when we applied the long-term strategy adoption rule compared to its short-term version (middle panel of Figure 2). We have received very similar results in all combinations of (a) other short- and long-term strategy adoption rule pairs; (b) a large number of other network topologies; (c) other games, such as the extended Prisoner's Dilemma or Hawk-Dove games. Long-term learning strategy adoption rules also promoted cooperation (albeit at lower efficiency than in case of complex network structures), when we used networks re-randomized after each play, or randomly picked agents (ESM1 Figures S1.3–S1.6). As a summary, we conclude that long-term strategy adoption rules ('learning' instead of simple imitation) allow a larger cooperation, but do not stabilize the cooperation-fluctuations inflicted by the different topologies of the underlying networks, which leaves the remarkable topological stability of the Q-learning strategy adoption rule still unexplained.

**Low level of randomness of the strategy adoption rules is needed to stabilize cooperation level in different network topologies**

Next we tested, if the innovative elements of the Q-learning strategy adoption rule may contribute to the stability of cooperation in various network topologies. For this, we constructed an 'innovative' version of the long-term version of the best-takes-over, 'non-innovative' strategy adoption rule by adding a low level of randomness instructing agents to follow the opposite of the selected neighbor's strategy with a pre-set $P_{innovation}$ probability (see Methods). Cooperation levels achieved by the innovative long-term best-takes-over strategy adoption rule are shown on the bottom panel of Figure 2. At temptation values smaller than T=3.8 the innovative long-term version of the best-takes over strategy adoption rule outperformed Q-learning, which resulted in a larger proportion of cooperating agents (cf. top and bottom panels of Figure 2). However, at high temptation values Q-learning proved to be more efficient in maintaining cooperation. Most importantly, cooperation levels in small-world and scale-free networks were much closer to each other, when using the long-term innovative strategy-adoption rule, than either the 'only long-term', or short-term versions of the same strategy adoption rule (Figure 2). At high temptation values cooperation levels of long-term innovative strategy adoption rules on small-world and scale-free networks were converging to each other and even to the cooperation level observed, when using the Q-learning strategy adoption rule. We have received very similar results in combinations of (a) other innovative short- and long-term strategy adoption rules; (b) a large number of other network topologies; (c) other games, such as the extended Prisoner's Dilemma or Hawk-Dove games (ESM1 Figures S1.7 and S1.8). According to the expectations [8], the stabilizing role of the randomness in the strategy adoption rules depended on the actual value of the pre-set $P_{innovation}$ probability, and showed an optimum at intermediary $P_{innovation}$ levels, where the actual value of optimal $P_{innovation}$ depended on the strategy adoption rule and network topology. The effect of changes in $P_{innovation}$ was much more pronounced in case of scale-free networks than at small-world networks, which is a rather plausible outcome, since the larger irregularity of scale-free networks makes the re-introduction of extinct strategies a lot more crucial (ESM1 Figure S1.8).

We have shown so far that long-term, learning strategy adoption rules help the development of cooperation, while 'innovative' strategy adoption rules make the cooperation level more independent from the actual network topology. Figure 3 illustrates how the cooperative network topologies were expanded, when we used long-term learning and 'innovative' versions of the best-takes-over strategy adoption rule as well as Q-learning at a high level of temptation, which made cooperation especially difficult. The application of the best-takes-over strategy adoption rule resulted in non-zero cooperation only sporadically. Cooperation levels using the long-term best-takes-over strategy adoption rule varied greatly, and still had several network configurations with zero cooperation. On the contrary, the two 'innovative' long-term learning strategy adoption rules had a much higher than zero cooperation in almost all networks tested, and the cooperation level remained fairly stable using a great variety of network topologies. This was especially true for Q-learning, which gave a stable level of cooperation even at regular networks (Figure 3), which result in a high instability of cooperation (see ESM1 Table S1.1). We have received very similar results in extended Prisoner's Dilemma and Hawk-Dove games (ESM1 Figures S1.9 and S1.10).

**Discussion**

As a summary, our simulations showed that long-term learning strategy adoption rules promote cooperation, while innovative elements make the appearance of cooperation less dependent from the actual network topology in two different games using a large number of network topologies in model networks. We must emphasize that the term 'learning' is used in our paper in the sense of the collection and use of information enriching and diversifying game strategy and behavior, and not in the restricted sense of imitation, or directed information-flow from a dominant source (the teacher) pauperizing the diversity of game strategies. The help of learning in promoting cooperation is already implicitly involved in the folk theorem, which opens the theoretical possibility for the emergence of cooperation at infinitely repeated games [3,20]. Learning, communication, negotiation, reputation-building mechanisms have all been shown to promote cooperation in various simulations as well as in games with groups of a variety of living organisms, including animals and humans (ESM1 Table S1.2). With the current work we have extended these findings showing that agents can markedly improve their cooperation, when they are allowed to consider long-term experiences either of their own (Q-learning) or their neighbors (other long-term strategies used), and this 'shadow of the past' [21] acts similarly at a great variety of network topologies.

We use the term 'innovation' in the sense of irregularities in the selection of adoption rules of game strategy. Therefore, 'innovation' may be caused by errors, mutations, mistakes, noise, randomness and temperature besides the *bona fide* innovation of conscious, intelligent agents. Our term, 'innovation' allows the change of the strategy adoption rules, therefore allows (increases) the evolvability [22] of our model system. Innovative strategies help to avoid 'herding', when agents start to use a uniform strategy and behavior forming synchronous clusters (ESM1 Figures S1.11, S1.12 and data not shown). Innovation increases game diversity and complexity, which, similarly to the stabilizing effect of weak links in a large variety of static networks, may significantly stabilize network dynamics (probably by helping the convergence of possible outcomes; [23]). Irregularities in network topology, noise, stochastic resonance, stochastic focusing and innovative strategies were shown to promote cooperation in various simulations as well as in games of primates and humans (ESM1 Table S1.3). However, the innovation-driven relative stabilization of cooperation in various network topologies is a novel finding reported here.

Cooperation helps the development of complex network structures [4,5,24]. Network dynamics and evolution lead to a large variety of link re-arrangements [25,26]. Network evolution is full of stochastic 'errors', and often results in the development of a higher average degree [25], which makes cooperation more difficult [15,16]. The highly similar cooperation levels of scale-free networks with different average degrees and of many other network topologies of model networks (Figure 3, ESM1 Figures S1.9 and S1.10) show that innovative long-term learning strategy adoption rules may provide a buffering safety-net to avoid the deleterious consequences of possible overshoots and errors in network development on cooperation. Our simulations showed (Figure 2, ESM1 Figures S1.2 and S1.6) that the help of innovative long-term learning is especially pronounced at conditions, where the relative cost of cooperation is the highest making cooperation most sensitive to the anomalies of network evolution [15].

This extreme situation is more easily reached, when the whole system becomes resource-poor, which makes all relative costs higher. Resource-poor networks develop a set of topological phase transitions in the direction of random → scale-free → star → fully connected subgraph topologies [27]. This further substantiates the importance of our findings that long-term, innovative learning allows a larger 'cooperation-compatible' window of these topologies, thus helps to avoid the decomposition of network structure in case of decreasing system resources due to e.g. an environmental stress. Further work is needed to show the validity of our findings in real-world networks as well as in combination with network evolution.

Our current work can be extended in a number of ways. The complexity of the game-sets and network topologies offers a great opportunity for a detailed equilibrium-analysis, similarly to that described by Goyal and Vega-Redondo [28]. The cited study [28] allows a choice of the interacting partners (an option denied in our model), which leads to another rich field of possible extensions, where the network topology is changing (evolving) during the games such as in the paper of Holme and Ghoshal [29]. Similarly, a detailed analysis of link rearrangement-induced perturbations, avalanches like in the paper of Ebel and Bornholdt [30] as well as exploration of a number of other topological re-arrangements would also significantly extend the current results. Such topology-changes may include

- hub-rewiring including the formation and resolution of 'rich-clubs', where hub-hub contacts are preferentially formed [31,32];
- emergence of modularity beyond to our data in ESM1 Figure S1.4;
- appearance and disappearance of bridge-elements between modules;
- changes of modular overlaps and module hierarchy, etc.

Tan [33] showed that cooperation helps faster learning. This, when combined with our current findings may lead to a self-amplifying cycle between cooperation and learning, where cooperation-induced learning promotes cooperation. Emerging cooperation alleviates a major obstacle to reach a higher level of network hierarchy and complexity [4]. In social networks learning establishes trust, empathy, reputation and embeddedness [34–37], and the benefits of learning by multiple generations are exemplified by the development of traditions, norms and laws. These give the members of the society further reasons for withholding their individual selfishness, thereby reaching a higher network complexity and stability. We believe that learning and innovation (in forms of repeated, interaction-driven, or random network remodeling steps, respectively or using the Baldwin-effect, see ESM1 Discussion) help the evolution of cooperation between agents other than human beings or animals, including proteins, cells or ecosystems [23,38], and were crucial in the development of multi-level, self-organizing, complex systems.

## Methods

**Games.** In both the Hawk-Dove and the Prisoner's Dilemma games, each agent had two choices: to cooperate or to defect. In the repeated, multi-agent Hawk-Dove game the benefit of defectors is higher than that of cooperators, when they are at low abundance, but falls below cooperator benefit, when defectors reach a critical abundance [12,13]. On the contrary, in the Prisoner's Dilemma game defection always has a fitness advantage over cooperation. The canonical parameter-set of the Prisoner's Dilemma game ( $R = 3, P = 1, S = 0$, the $T$, temptation value varies between 3 to 6; 3 is not included; where R is the reward for mutual cooperation, P is the punishment for mutual defection, S and T are the payoffs for the cooperator and defector, respectively, when meeting each other) restricts cooperation more, than the parameter set of the extended (also called as 'weak') Prisoner's Dilemma game ($R=1, P=0, S=0$ with $T$ values ranging from 1 to 2; [11–13]). (When we tried the parameter set of $R=1, P=0.2, S=0.1$ with T values ranging from 1.0 to 2.0, we have received very similar results; data not shown.)

In the Hawk-Dove games (or in the conceptually identical Snowdrift and Chicken games [13,39,40]) each agent had two choices: to defect (to be a hawk) or to cooperate (to be a dove). When a hawk met a dove, the hawk gained $G$ benefits, whereas the payoff for the dove was zero. Two hawks suffered a $(G-C)/2$ cost each upon encounter, where $C>G$ was the cost of their fight. When two doves met, the benefit for each dove was $G/2$. If not otherwise stated, the cost of injury (C, when a hawk met a hawk) was set to 1. The value of G varied from 0 to 1 with the increments of 0.1. If we want to compare the above, usually applied nomenclature of the Hawk-Dove games with that of the Prisoner's Dilemma games, R=G/2, P=(G-C)/2, S=0 and T=G.

In Hawk-Dove games T>R>S>P, in the extended (also called 'weak') Prisoner's Dilemma game T≥R>P≥S, while in the canonical Prisoner's Dilemma game T>R>P>S. This makes the following order of games from less to more stringent general conditions allowing less and less cooperation: Hawk-Dove game > extended Prisoner's Dilemma game > canonical Prisoner's Dilemma game. Due to this general order, we showed the results of the canonical Prisoner's Dilemma game in the main text, and inserted the results of the two other games to the Electronic Supplementary Material 1 (ESM1).

In our simulations each node in the network was an agent, and the agent could interact only with its direct neighbors. Agents remained at the same position throughout all rounds of the repeated games, and they were neither exchanged, nor allowed to migrate. If not otherwise stated, games started with an equal number of randomly mixed defectors and cooperators (hawks and doves in the Hawk-Dove game), and were run for 5,000 rounds (time steps). The payoff for each agent in each round of play was the average of the payoffs it received by playing with all its neighbors in the current round. In our long-term learning strategy adoption rules introduced below, the accumulative payoff means the accumulation of the average payoffs an agent gets in each round of play. Average payoff smoothes out possible differences in the degrees of agents, and in several aspects may simulate real-world situations better than non-averaged payoff, since in real-world situations agents usually have to observe a cost of maintaining a contact with their neighbors [39–41]. Moreover, average payoff helps the convergence of cooperation levels as the rounds of the game (time steps) proceed, what we indeed observed in most of the cases (with a few exceptions noted in the text), and helps to avoid 'late-conversions' occurring mostly in scale-free networks after 10,000 or more time steps using non-averaged payoffs. With this method it was enough to calculate the proportion of cooperators as the average ratio of cooperators of the last 10 rounds of the game (if not otherwise stated) for 100 independent runs.

**Strategy adoption rules.** In Prisoner's Dilemma and Hawk-Dove games our agents followed three imitation-type, short-term strategy adoption rules, the 'pair-wise comparison dynamics' (also called as 'replicator dynamics'), 'proportional updating' and 'best-takes-over' (also called as 'imitation of the best') strategy adoption rules [13]. We call these rules strategy adoption rules and not evolution rules to avoid the mis-interpretation of our games as cellular automata-type games, where agents are replaced time-to-time. In our games no replacement took place, therefore these games were not evolutionary games in this strict sense. All strategy adoption rules had synchronous update, meaning that in each round of play the update took place after each agent had played with all their neighbors. To avoid the expansion of parameters with the differential placements of various agents in complex network structures all agents used the same strategy adoption rule in the agent-network. In the three strategy adoption rules we applied initially ('best-takes-over', 'pair-wise comparison dynamics' and 'proportional updating') all agents were myopic, and made their decisions based on the average payoffs gained in the previous round.

**Pair-wise comparison dynamics strategy adoption rule.** In the 'pair-wise comparison dynamics' strategy adoption rule [13] for any agent $i$, a neighboring agent $j$ was selected randomly, and agent $i$ used the strategy of agent $j$ with a probability of $p_i$. In our experiments the probability was determined as

$$p_i = f(G_i - G_j) = \begin{cases} \dfrac{G_j - G_i}{d_{max}} & \text{if } G_j - G_i > 0 \\ 0 & \text{otherwise} \end{cases},$$

where $d_{max} = (G+C)/2$ (for Hawk-Dove games) or $d_{max} = \max(T,R)$ (for Prisoner's Dilemma games), which was the largest gap of gain between two agents in one round of play. $G_i$ and $G_j$ were the average payoffs received by agent $i$ and $j$ respectively in the current round of play.

**Proportional updating strategy adoption rule.** For the 'proportional updating' strategy adoption rule [13] agent $i$ and all its neighbors competed for the strategy of agent $i$ with the probability $p_i$, which was determined as $p_i = \dfrac{G_i}{\sum_n G_n}$, $i,n \in \{N(i) \cup i\}$ where $N(i)$ was the neighborhood of agent $i$ and $G_i$ was the average payoff received by agent $i$ in the current round of play. Since $p$ is a probability, $C$ was added to each $G_i$ to avoid negative values. For Prisoner's Dilemma games, because the reward for an agent is always greater than or equal to zero, there was no need to increase the value of $G_i$.

**Best-takes-over strategy adoption rule.** In the 'best-takes-over' strategy adoption rule (also called as imitation of the best strategy adoption rule, [13]) agent $i$ adopted the strategy of that agent selected from $i$ and its neighbors, who had the highest average payoff in the last round of play.

**Q-learning strategy adoption rule.** As a reinforcement learning [19] strategy adoption rule, we used Q-learning [18], where agents learned an optimal strategy maximizing their total discounted expected reward in the repeated game. In Q-learning we assumed that the environment constituted a discrete Markov process with finite states. An agent chose action $a_t$ from a finite collection of actions at time step, $t$. The state of the environment changed from state $s_t$ to $s_{t+1}$ after the action of the agent, and the agent received the reward $r_t$ at the same time. The probability of state transition from $s_t$ to $s_{t+1}$ when the agent chose action $a_t$ was

$$prob[s = s_{t+1} \mid s_t, a_t] = P[s_t, a_t, s_{t+1}].$$

The task of the agent was to learn the optimal strategy to maximize the total discounted expected reward. The discounted reward meant that the rewards received by the agent in the future were worth less than that received in the current round. Under a policy $\pi$ denoting how the agent selected the action at its actual state and reward, the value of state, $s_t$ was

$$V^\pi(s_t) = R(\pi(s_t)) + \gamma \sum_{s_{t+1} \in S} P[s_t, a_t, s_{t+1}] V^\pi(s_{t+1}),$$

where $R(\pi(s_t))$ is the expected reward of state $s_t$ under policy $\pi$ and $\gamma$ ($0 < \gamma < 1$) is the discount factor.

The theory of Dynamic Programming [19] guarantees that there is at least one optimal stationary policy, $\pi^*$, which can be written as

$$V^*(s_t) = V^{\pi^*}(s_t) = \max_{a_t \in A} \left\{ R(\pi(s_t)) + \gamma \sum_{s_{t+1} \in S} P[s_t, a_t, s_{t+1}] V^{\pi^*}(s_{t+1}) \right\}.$$

The task of Q-learning was to learn the optimal policy, $\pi$, when the initial conditions of both the reward function and transition probabilities were unknown. If the environment model (reward model and transition probabilities of states) is known, then the above problem can be solved by using Dynamic Programming. Watkins and Dayan [18] introduced Q-learning as incremental Dynamic Programming.

The idea of Q-learning is to optimize a Q-function, which can be calculated iteratively without the estimate of environment model. For this having a policy, $\pi$, we defined the Q-value as:

$$Q(s,a) = R(\pi(s)) + \gamma \sum_{s' \in S} P[s,a,s'] V^{\pi}(s').$$

Q-learning consisted of a sequence of distinct stages or episodes. The Q value of state-action pair $(s_t, a_t)$ can be learned through the following iterative method:

$$Q_t(s_t, a_t) = (1-\alpha_t) Q_{t-1}(s_t, a_t) + \alpha_t [r_t + \gamma V_{t-1}(s_{t+1})],$$

where $V_{t-1}(s_{t+1}) = \max_{a \in A}\{Q_{t-1}(s_{t+1}, a)\}$ and $\alpha_t$ controls the learning and convergence speed of Q-learning.

In repeated multi-agent games, the state of each agent was affected by the states of its direct neighbors. Those neighbors constituted the environment of the agent. The reward of the agent $i$ after taking action $a_t(i)$ was defined as:

$$r_t(i) = \frac{1}{k_i} \sum_{j \in N(i)} S_t^T(i) M S_t(j),$$

where $M$ was the payoff matrix, $S_t(i)$ was a column vector indicating the state of agent $i$ at round $t$, $k_i$ was the number of neighbors of agent $i$ and $N(i)$ was the set contains all the direct neighbors of agent $i$. The values of elements of $S_t(i)$ were 0 or 1 and 1 indicated that agent $i$ was in the corresponding state. In such a repeated multi-agent game, Q-learning meant that each agent tried to optimize its total discounted expected reward in the repeated game. The optimal strategy was approximated by an iterative annealing process. For this for each agent, the selection probability (Boltzmann-probability) of action $a_i$ at time step $t$ was defined as

$$prob(a_i) = \frac{e^{Q(s_t, a_i)/T}}{\sum_{a_k \in A} e^{Q(s_t, a_k)/T}},$$

where $T$ was the annealing temperature. In our experiments we selected the discount factor, $\gamma_t = 0.5$, since in the initial experiments we found that this value is helpful to achieve high levels of cooperation. The initial annealing temperature was set to 100 in Hawk-Dove and extended Prisoner's Dilemma games, while it was raised to 10,000 in canonical Prisoner's Dilemma games to extend the annealing process [42]. In all cases the annealing temperature was decreased gradually by being divided by $t$ in each round of the game till it reached a low bound of 0.001. In order to control the convergence speed of Q-learning, $\alpha = 1/(1+TimesVisited(s,a))$ where $TimesVisited(s,a)$ was the number of times that the state-action pair $(s,a)$ had been visited at time step $t$. In this way $\alpha$ decreased gradually with the time.

**Long-term learning and innovative strategy adoption rules.** Long-term learning strategy adoption rules were generated by considering the accumulative average payoffs instead of instantaneous average rewards in the update progress during each round of play for all strategy adoption rules used. In both short term and long-term innovative strategy adoption rules, agent $i$ used the opposite strategy of the selected neighbor (for proportional updating and best-takes-over strategy adoption rules, the neighborhood included agent $i$ itself) in the last round of play with probability of $P_{innovation}$, which was 0.0001 in case of Hawk-Dove and extended Prisoner's Dilemma games, while 0.0002 in case of canonical Prisoner's Dilemma games, if not otherwise stated (like in the legend of ESM1 Figure S1.8). In innovative strategy adoption rules agent $i$ adopted the strategy of the selected neighbor with a probability of $1 - P_{innovation}$.

**Network construction.** In our work we used a set of widely adopted model networks to simulate the complexity of real-world situations. Generation of the Watts-Strogatz-type small-world model network [17] was modified according to Tomassini *et al.* [13] to avoid the heterogeneity in node degrees, which arose during the Watts-Strogatz-type rewiring process changing the regular lattice to a small-world network. Such heterogeneity was shown to have a rather big influence on the level of cooperation [13,43]. At the generation of the Barabasi-Albert-type scale-free network [44], we started from an initial fully connected graph of 'm' nodes (where 'm' ranged from 1 to 7), and added the new nodes with 'm' novel links as specified at the individual Figure legends. In the modular networks described by Girvan and Newman [45] each network had a scale-free degree distribution, contained 128 nodes, and was divided into 4 communities. The average degree was 16. Modularity (community structure) was gradually decreased at 'levels' 1, 5, 10 and 16, where 'level 1' meant that for each node in the network, the expected number of links between a node and the nodes which were in other communities was 1 (e.g. low compared to the average degree of 16). With increasing 'level' the community structure gradually decreased.

**Network visualization.** At the visualization the coordinates of the small-world networks with a rewiring probability of p=0.01 were used for the p=0.04 networks to avoid the individual variations of the Pajek-figures [46] and to help direct comparison. With 15x15 agents the final representations of cooperators showed a moderate variability. This was almost negligible, when 50x50 agents were used (data not shown). However, 15x15 agents gave a better visual image than the crowded, bulky 50x50 version. Therefore, we opted to include this variant to Figure 1. We have selected those figures from the results of 15x15 agent games, which best represented the 50x50 versions.

## Acknowledgments

Work in the authors' laboratory was supported by research grants from the EU (FP6-518230), Hungarian Science Foundation (OTKA- K69105), the National Natural Science Foundation of China (NSFC-60721003) and the Hungarian National Research Initiative (NKFP-1A/056/2004 and KKK-0015/3.0). The useful comments of our Editor, Enrico Scalas, referees, including Michael König and Bence Toth, as well as Robert Axelrod, János Kertész, István A. Kovács, Robin Palotai, György Szabó, Attila Szolnoki, Tamás Vicsek and members of the LINK-Group (www.linkgroup.hu) are gratefully acknowledged.

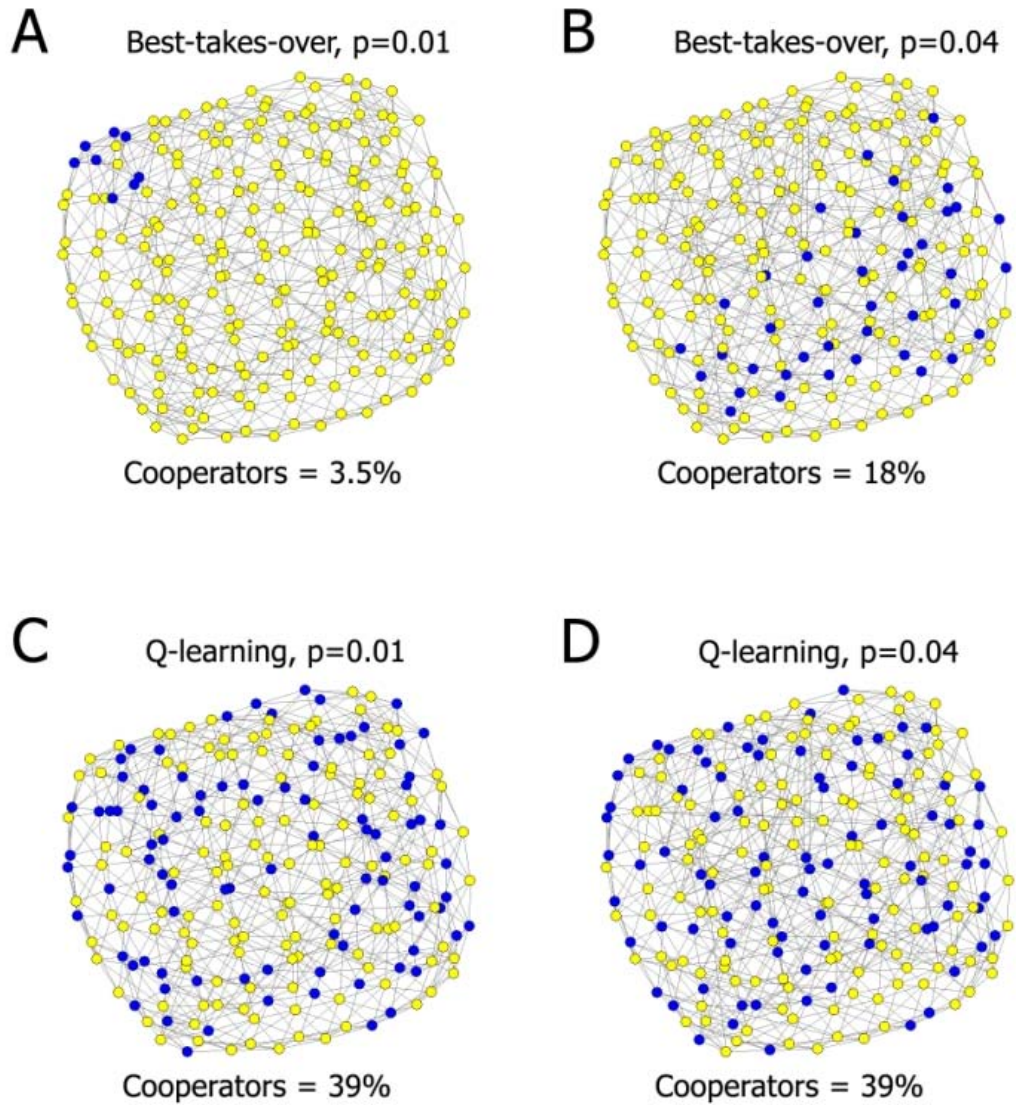

**Figure 1.** A long-term learning adoption rule, Q-learning improves and stabilizes cooperation of agents forming various small-world networks in Prisoner's Dilemma games

The modified Watts-Strogatz small-world network was built on a 15 x 15 lattice, where each node was connected to its eight nearest neighbors. The rewiring probabilities of the links placed originally on a regular lattice were 0.01 (left panels) and 0.04 (right panels), respectively. For the description of the canonical repeated Prisoner's Dilemma game, as well as the best-takes-over (top panels) and Q-learning (bottom panels) strategy adoption rules see Methods and the ESM1. The temptation level, T was 3.6. Networks showing the last round of 5,000 plays were visualized using the Kamada-Kawai algorithm of the Pajek program [46]. Dark blue dots and diamonds correspond to cooperators and defectors, respectively. The Figure shows that both the extent and distribution of cooperators vary, when using the best-takes-over strategy adoption rule (see top panels), while they are rather stable with the Q-learning strategy update rule (see bottom panels).

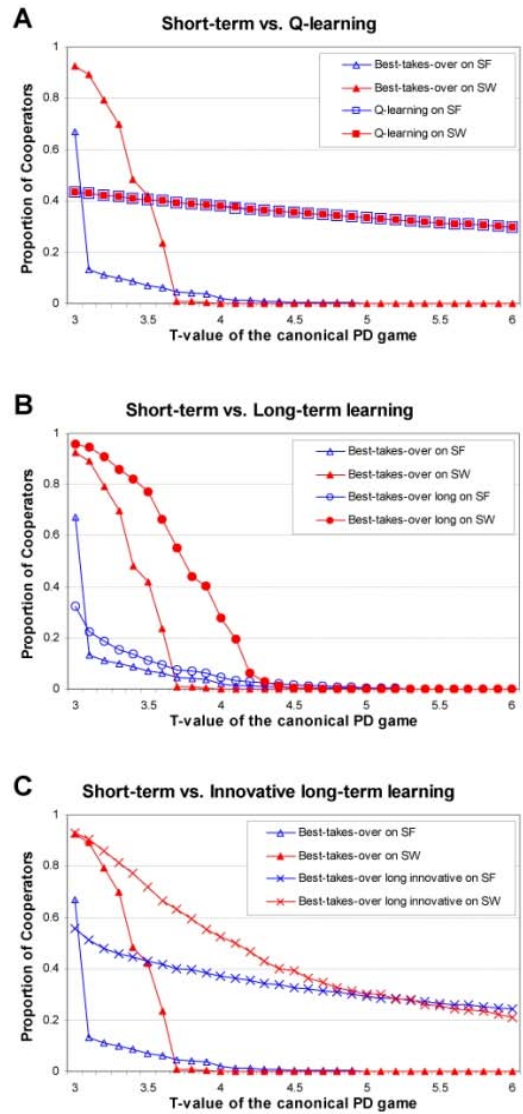

**Figure 2.** Long-term learning elements of strategy update rules help, while a low level of randomness relatively stabilizes cooperation in Prisoner's Dilemma games played on various networks

Small-world (SW, filled, red symbols) networks were built as described in the legend of Figure 1. The Barabasi-Albert-type scale-free networks (SF, open, blue symbols) contained 2,500 nodes, where at each construction step a new node was added with 3 new links attached to the existing nodes. For the description of the canonical repeated Prisoner's Dilemma game, as well as that of the best-takes-over (triangles, all panels), the Q-learning (rectangles, top panel) the best-takes-over long (circles, middle panel), and the best-takes-over long innovative (crosses, $P_{innovation}$ = 0.0002, bottom panel) strategy adoption rules, see Methods and the ESM1. For each strategy adoption rules and $T$ temptation values 100 random runs of 5,000 time steps were executed. The figure shows that long-term, 'learning-type' elements of strategy update rules help cooperation in Prisoner's Dilemma games played on various networks. A low level of randomness (also called as 'innovation' in this paper) brings the level of cooperation closer in different network topologies.

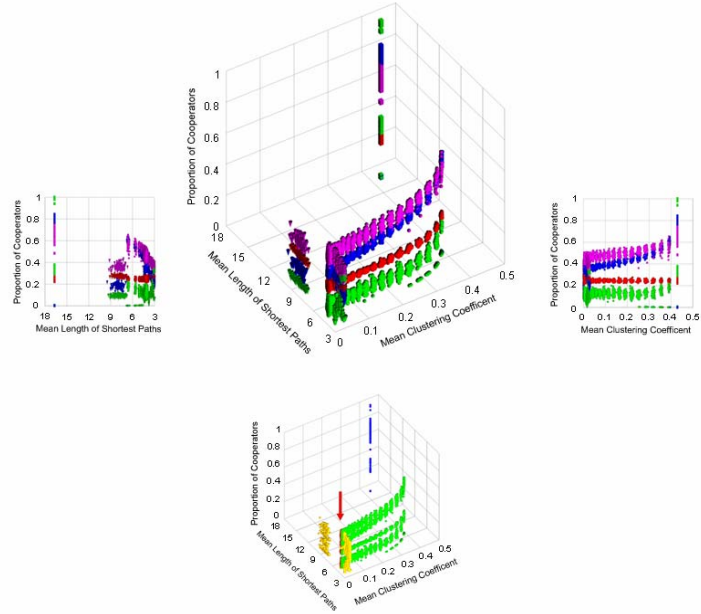

**Figure 3.** Long-term learning and innovative elements of strategy adoption rules, when applied together allow cooperation in a large number of model networks.

(Top middle panel) The small-world (spheres) and scale-free (cones) model networks were built as described in the legends of Figures 1 and 2. The rewiring probability, $p$ of the links of the original regular lattices giving small-world networks was increased from 0 to 1 with 0.05 increments, the number of edges linking each new node to former nodes in scale-free networks was varied from 1 to 7, and the means of shortest path-lengths and clustering coefficients were calculated for each network. Cubes and cylinders denote regular ($p = 0$) and random ($p = 1.0$) extremes of the small-world networks, respectively. For the description of the canonical repeated Prisoner's Dilemma game, as well as the best-takes-over (green symbols); long-term learning best-takes-over (blue symbols); long-term learning innovative best-takes-over (magenta symbols) and Q-learning (red symbols) strategy adoption rules used, see Methods and the ESM1. For each network 100 random runs of 5,000 time steps were executed at a fixed $T$ value of 3.5.

(Left and right panels) 2D side views of the 3D top middle panel showing the proportion of cooperators as the function of the mean length of shortest paths or the mean clustering coefficient, respectively.

(Bottom middle panel) Color-coded illustration of the various network topologies used on the top middle panel. Here the same simulations are shown as on the top middle panel with a different color-code emphasizing the different network topologies. The various networks are represented by the following colors: regular networks – blue; small-world networks – green; scale-free networks – yellow; random networks – red (from the angle of the figure the random networks are behind some of the small-world networks and, therefore are highlighted with a red arrow to make there identification easier).

The top middle panel and its side views show that the best-takes-over strategy adoption rule (green symbols) at this high temptation level results in a zero (or close-to-zero) cooperation. As opposed to this, the long-term best-takes-over strategy adoption rule (blue symbols) raise the level of cooperation significantly above zero, but the individual values vary greatly at the different network topologies. When the long-term strategy adoption rule is combined with a low level of randomness (magenta symbols) the cooperation level stays in most cases uniformly and its variation becomes high greatly diminished. Q-learning stabilizes cooperation further even at regular networks, which otherwise give an extremely variable outcome.

Electronic Supplementary Material (ESM) of

# Learning and innovative elements of strategy adoption rules expand cooperative network topologies

Shijun Wang, Máté S. Szalay, Changshui Zhang & Peter Csermely*

*To whom correspondence should be addressed. E-mail: csermely@puskin.sote.hu

**Content of this electronic supplementary material**

# Supplementary Text

## Supplementary Results

Similarly to the case shown in Figures 1 and 2 for the best-takes-over strategy adoption rule in canonical Prisoner's Dilemma games, the three short-term strategy adoption rules (pair-wise comparison dynamics, proportional updating and best-takes-over) resulted in a rather remarkable variation of cooperator levels in Hawk-Dove games when using large number of small-world and scale-free model networks (Figures S1.1 and S1.2). For the description of game types, strategy adoption rules and model networks see Methods and refs. [1-11].

At Figure S1.1 the m=1 scale-free networks display an irregular 'phase-transition'-like phenomenon, which is most pronounced at the proportional updating strategy adoption rule but leads to a faster decay of cooperation at all short-term strategy adoption rules tested. At the construction of these m=1 scale-free networks the novel nodes are linked to the existing network with a single link only, which results in a tree-like final topology. Due to the especially large wiring-irregularity of these networks (as compared to the similarly scale-free, but more 'cross-linked' networks, where the new nodes are joined with more than one links to the existing network) a gradual change in the payoff values makes a more rapid disappearance of cooperation. At panel E of Figure S1.1 a non-monotonic behavior of p=0 networks is observed. This is derived from the extreme sensitivity of these p=0 regular networks on initial conditions, strategy update rules, etc (see references listed in Table S1.1).

Both Q-learning and the long-term versions of all three strategy adoption rules above outperformed the short-term variants resulting in a higher proportion of cooperators in Hawk-Dove games on small-world and scale-free model networks especially at high cooperation costs (Figures S1.2*A*, S1.2*B* and S1.3). Long-term strategy adoption rules (including Q-learning) were also more efficient inducers of cooperation even at high costs in modular networks (Figure S1.7). Moreover, long-term strategy adaption rules maintained cooperation even in randomly mixed populations as well as in repeatedly re-randomized networks (Figure S1.5). Interestingly, long-term strategy adoption rules (especially the long-term version of the best-takes-over strategy adoption rule) resulted in an extended range of all-cooperator outcomes in Hawk-Dove games (Figures S1.3–S1.5 and S1.7). Finally, long-term strategy adoption rules helped cooperation in canonical and extended Prisoner's Dilemma games in case of all three strategy adoption rules tried (Figure S1.6).

While short- and long-term strategy adoption rules resulted in a remarkable variation of the cooperation level in a large variety of random, regular, small-world, scale-free and modular networks in Hawk-Dove and both canonical and extended Prisoners' Dilemma games (Figures S1.1–S1.6), Q-learning induced a surprising stability of cooperation levels in all the above circumstances (Figures S1.2–S1.6). Interestingly, but expectedly, Q-learning also stabilized final cooperation levels, when games were started from a different ratio of cooperators (ranging from 10% to 90%) than the usual 50% (data not shown). When we introduced innovativity to long-term strategy adoption rules in Hawk-Dove games (for the description of these innovative strategy adoption rules see Methods) similarly to that shown for the canonical Prisoner's Dilemma game on Figure 2, cooperation levels were closer to each other in small-world and scale-free networks than their similarity observed when using only long-term, but not innovative strategy adoption rules (Figure S1.7). Importantly, innovativity alone, when applied to the best-takes-over short-term strategy adoption rule could also stabilize cooperation levels in small-world and scale-free networks (Figure S1.7*C*). When we compared different levels of innovation by changing the value of $P_{innovation}$ in our simulations (Figure S1.8), an intermediary level of innovation was proved to be optimal for the stabilization of cooperation in small-world and scale-free networks. Scale-free networks and Prisoner's Dilemma game were more sensitive to higher innovation levels than small-world networks or Hawk-Dove games, respectively (Figure S1.8). Summarizing our results, Figures S1.9 and S1.10 show that similarly to canonical Prisoner's Dilemma games (Figure 3), both in Hawk-Dove games (Figure S1.9) and extended Prisoner's Dilemma games (Figure S1.10) long-term strategy adoption rules and innovation (including Q-learning) resulted in a stable non-zero cooperation in a large variety of network topologies in combination only.

Figure S1.11 shows the distribution of hawks (blue dots) and doves (orange dots) at the last round of a repeated Q-learning game on small-world (Figure S1.11$A$ and S1.11$B$) or scale-free networks (Figure S1.11$C$ and S1.11$D$) at low (Figure S1.11$A$ and S1.11$C$) and high (Figure S1.11$B$ and S1.11$D$) relative gain/cost ($G$) values. Under these conditions both hawks and doves remained isolated (see arrows). On the contrary, when Hawk-Dove games were played with any of the three short-term, non-innovative strategy adoption rules doves, but even hawks showed a tendency to form networks (Figure S1.12 and data not shown). This effect was especially pronounced for doves in both small-world and scale-free networks, as well as for hawks in small-world networks, and present, but not always that strong for hawks in scale-free networks, where hawks remained more isolated in all configurations. Interestingly, the proportional updating strategy adoption rule quite often showed an extreme behavior, when in the last round of the play all agents were either doves or hawks. This behavior was less pronounced with a larger number (2,500) of players. All the above findings were similarly observed in extended Prisoner's Dilemma games (data not shown).

## Supplementary Discussion

Explaining cooperation has been a perennial challenge in a large section of scientific disciplines. The major finding of our work is that learning and innovation extend network topologies enabling cooperative behavior in the Hawk-Dove (Figures S1.1–S1.5 and S1.7–S1.9, S1.11, S1.12) and even in the more stringent Prisoner's Dilemma games (Figures 1–3, S1.6, S1.8 and S1.10). The meaning of 'learning' is extended here from the restricted sense of imitation or learning from a teacher. Learning is used in this paper to denote all types of information collection and processing to influence game strategy and behavior. Therefore, learning here includes communication, negotiation, memory and various reputation building mechanisms. Learning makes life easier, since instead of the cognitive burden to foresee and predict the 'shadow of the future' [4–6] learning allows to count on the 'shadow of the past', the experiences and information obtained on ourselves and/or other agents [12]. Likewise to our understanding of learning, the meaning of 'innovation' is extended here from the restricted sense of innovation by conscious, intelligent agents. Innovation is used in this paper to denote all irregularities in the strategy adoption process of the game. Therefore, innovation here includes errors, mutations, mistakes, noise, randomness and increased temperature besides conscious changes in game strategy adoption rules.

In the Supplementary Discussion, first we summarize the effects of network topology on cooperative behavior, then discuss the previous knowledge on the help of cooperation by learning and innovation, and, finally, we compare our findings with existing data in the literature and show their novelty and implications.

**Effect of network topology on cooperation.** Cooperation is not an evolutionary stable strategy [13], since in the well-mixed case, and even in simple spatial arrangements it is outcompeted by defectors. As it is clear from the data summarized in Table S1.1, the emergence of cooperation requires an extensive spatial segregation of players helping cooperative communities to develop, survive and propagate. Cooperation in repeated multi-agent games is very senitive to network topology. Cooperation becomes hindered, if the network gets over-connected [14–16]. On the contrary, high clustering [17,18], the development of fully connected cliques (especially overlapping triangles) and rather isolated communities [14,18] usually help cooperation. Heterogeneity of small-worlds and, especially, networks with scale-free degree distribution can establish cooperation even in cases, when the costs of cooperation become exceedingly high.

However, in most spatial arrangements cooperation is rather sensitive to the strategy adoption rules of the agents, and especially to the strategy adoption rules of those agents, which are hubs, or by any other means have an influential position in the network. Moreover, minor changes in the average degree, actual degree, shortests paths, clustering coefficients or assortativity of network topology may induce a profound change in the cooperation level. Since real world networks may have rather abrupt changes in their topologies [17,20–26], it is highly important to maintain cooperation during network evolution.

**Effect of learning on cooperation.** From the data of Table S1.2 it is clear that learning generally helps cooperation. Cooperation can already be helped by a repeated play, assuming 'learning' even among spatially disorganized players. Memory-less or low memory strategy adoption rules do not promote cooperation efficiently. In contrast, high-memory and complex negotiation and reputation-building mechanisms (requiring the learning, conceptualization and memory of a whole database of past behaviors, rules and motives) can enormously enhance cooperation making it almost inevitable. As a summary, in the competitive world of games, it pays to learn to achieve cooperation. However, it is not helpful to know too much: if the ranges of learning and the actual games differ too much, cooperation becomes impossible [18].

Learning requires a well-developed memory and complex signaling mechanisms, which are costly. This helps the selection process in evolution [13], since 'high-quality' individuals can afford the luxury of both the extensive memory and costly signaling [27]. However, cooperation is rather widespread among bacteria, where even the 'top-quality individuals' do not have the extensive memory mentioned above. Here 'learning' is achieved by the fast succession of multiple generations. The Baldwin-effect describing the genetic (or epigenetic) fixation of those behavioral traits, which were benefitial for the individuals, may significantly promote the development of bacterial cooperation and the establishment of biofilms [28–32]. Genetically 'imprinted' aids of cooperation are also typical in higher organisms including humans. The emotional reward of cooperation uncovered by a special activation of the amygdalia region of our brains [33] may be one of the genetically stabilized mechanisms, which help the extraordinary level of human cooperation besides the complex cognitive functions, language and other determinants of human behavior.

**Effect of randomness ('innovation') on cooperation.** From the data of Table S1.3 it is clear that a moderate amount of randomness, 'innovation' generally helps cooperation. Many of the above learning mechanisms imply sudden changes, innovations. Bacteria need a whole set of mutations for interspecies communication (such as quorum sensing), which adapt individual organisms to the needs of cooperation in biofilms or symbiotic associations. The improved innovation in the behavior of primates and humans during games has been well documented [34–36].

An appropriate level of innovation rescues the spatial assembly of players from deadlocks, and accelerates the development of cooperation [18]. Many times noise acts in a stochastic resonance-like fashion, enabling cooperation even in cases, when cooperation could not develop in a zero-noise situation [37,38]. As a special example, the development of cooperation between members of a spatial array of oscillators (called synchrony) is grossly aided by noise [39]. Egalitarian motives also introduce innovative elements to strategy selection helping the development of cooperation [40].

However, innovation serves the development of cooperation best, if it remains a luxurious, rare event of development. Continuous 'innovations' make the system so noisy, that it looses all the benefits of learning and spatial organization and reaches the mean-field limit of randomly selected agents with random strategy adoption rules (Table S1.3).

**Comparison and novelty of our findings.** In Hawk-Dove games on modified Watts-Strogatz-type small-world [2,9] and Barabasi-Albert-type [10] scale-free model networks we obtained very similar results of cooperation levels in all synchronously updated pair-wise comparison dynamics, proportional updating and best-takes-over strategy adoption rules to those of Tomassini et al. [2,3]. The success of our various 'long-term' strategy adoption rules to promote cooperation is in agreement with the success of pair-wise comparison dynamics and best-takes-over strategy adoption rules with accumulated payoffs on scale-free networks [1,3].

On the contrary to Hawk-Dove games, in the Prisoner's Dilemma game defection always has a fitness advantage over cooperation, which makes the achievement of substantial cooperation levels even more difficult. In the extended Prisoner's Dilemma games on scale-free networks [10] we obtained very similar results of cooperation levels using synchronously updated pair-wise comparison dynamics and best-takes-over strategy adoption rules to those of Tomassini et al. [3]. Similarly to the Hawk-Dove game with the extended Prisoner's Dilemma game our results with various 'long-term' strategy adoption rules on scale-free networks are in agreement with those of pair-wise comparison dynamics and best-takes-over strategy adoption rules using accumulated payoffs [1,3].

We have to note that the definition of pair-wise comparison dynamics strategy adoption rule was slightly different here, than in previous papers, and on the contrary to the non-averaged payoffs used previously, we used average payoffs [1–3], which allows only a rough comparison of these results to those obtained before, and resulted in a lower level of cooperation than that of e.g. ref. [1]. The reason we used average payoff was that this made the final level of cooperators more stable at scale-free networks even after the first 5,000 rounds of the play (data not shown). When we used non-averaged payoffs in the extended Prisoner's Dilemma game with 100,000 rounds of play, we re-gained the cooperation levels of ref. [1] at scale-free networks (m=4, data not shown). The additional papers on the subject used differently designed small-world networks or different strategy adoption rules, and therefore can not be directly compared with the current data. It is worth to mention that none of the previous papers describing multi-agent games on various networks [1–3] used the canonical Prisoner's Dilemma game, which was used obtaining our data in the main text, and which gives the most stringent condition for the development of cooperation.

As a summary, our work significantly extended earlier findings, and showed that the introduction of learning and innovation to game strategy adoption rules helps the development of cooperation of agents situated in a large variety of network topologies. Moreover, we showed that learning and innovation help cooperation separately, but act synergistically, if introduced together especially in the complex form of the reinforcement learning, Q-learning.

**Interactions of learning and innovations, conclusions.** Real complexity and excitement of games needs both learning and innovation. In Daytona-type car races skilled drivers use a number of reputation-building and negotiation mechanisms, and by continuously bringing novel innovations to their strategies, skilfully navigate between at least four types of games [41].

Noise is usually regarded to disturb the development of cooperation. Importantly, complex learning strategies can actually utilize noise to drive them to a higher level of cooperation. Noise may act as in the well-known cases of stochastic resonance, or stochastic focusing (with extrinsic and intrinsic noise, respectively) enabling cooperation even in cases, when it could not develop without noise. In a similar fashion, mistakes increase the efficacy of learning [37,38,42]. Additional noise greatly helps the optimization in the simulated annealing process [43–45].

Noise not only can extend the range of cooperation to regions, where the current level of learning would not be sufficient to achieve it, but extra learning can also 'buffer' an increased level of noise [19]. Thus, learning and innovation act side-by-side and – in gross terms – correct the deficiencies of the other. Learning and innovation also cooperate in the Baldwin effect, where beneficial innovations (in the form of mutations) are selected by the inter-generational 'meta-learning' process of evolution [28–32]. Mutual learning not only makes innovation tolerable, but also provokes a higher level of innovation to surpass the other agent [36].

Our work added the important point to this emerging picture that the cooperation between learning and innovation to achieve cooperation also works in the extension and buffering of those network configurations, where cooperation becomes possible.

## Supplementary References

# Supplementary Tables

**Table S1.1.** Effect of network topology on cooperation

| Network topology | Effect on cooperation | Games; strategy adoption rules | Agents (players) | References |
|---|---|---|---|---|
| **Lattice** | **Sensitive to strategy adoption rules and topology** (cooperation level is very sensitive on strategy adoption rules, high degree *inhibits* cooperation) | HD, PD[a] | Simulation | 14, 46–48 |
| **Lattice with dilution** (with empty spaces) | **Helps** (localized groups of cooperators emerge better) | PD | Simulation | 49 |
| **Lattice with hierarchical layers** | **Helps** (at top level, if the number of levels is lower than 4; in middle layers otherwise) | PD | Simulation | 50, 51 |
| **Regular random graphs** | **Sensitive to topology** (triangles help, loops>3 and high degree *inhibit* cooperation) | PD | Simulation | 14, 52 |
| **Random graphs** | **Sensitive to topology, long-lasting avalanches may develop** (high degree *inhibits* cooperation) | PD | Simulation | 14, 15, 53 |
| **Small-world (Watts-Strogatz-type)** | **Mostly helps** (helps the spread of cooperation + introduces heterogeneity to stabilize it, high degree *inhibits* cooperation) | PD | Simulation | 14, 15, 54, 55 |
| **Small-world (randomly replaced edges)** | **Sensitive to strategy adoption rules** (very sensitive to the applied strategy adoption rules) | HD | Simulation | 2, 3 |
| **Small-world (Watts-Strogatz-type) with an influential node** | **Destabilizes** (the central node is very sensitive for attacks by defectors) | PD | Simulation | 56 |
| **Homogenous small-world** (degree is kept identical) | **Sensitive to topology and temptation level** (at small temptation helps the attack of defectors via shortcuts, helps at high temptation) | PD | Simulation | 54, 55 |
| **Scale-free (Barabasi-Albert-type)** | **Sensitive to to strategy adoption rules** (hubs stabilize cooperation but make it vulnerable to targeted attacks, clusters and loops help, sensitive to strategy adoption rules, high degree *inhibits* cooperation) | HD, PD; pair-wise comparison dynamics, imitation of the best | Simulation | 1, 3, 14–16, 57–59 |

**Table S1.1.** Effect of network topology on cooperation (continued)

| Network topology | Effect on cooperation | Games; strategy adoption rules | Agents (players) | References |
|---|---|---|---|---|
| **Scale-free with hierarchy** (Ravasz-Barabasi-type hierarchy) | **Inhibits** (makes it very sensitive for the attack of defectors) | PD | Simulation | 50 |
| **Scale-free with communities** | **Helps** (isolated communities help intra-community cooperation) | PD | Simulation | 16 |
| **Real world networks** | **Generally helps** (small-worlds and hierarchy help cooperation) | PD | Internet communities, emails, karate club | 60 |
| **Dynamic** (evolves during the game) | **Generally helps** (a small-world and hierarchy develops, which stabilizes cooperation, a slower reaction to new information is beneficial) | PD | Simulation | 17, 21–26 |

[a]HD = Hawk-Dove (Snowdrift, Chicken) game; PD = Prisoner's Dilemma game (please note that in this supplementary table we did not discriminate between conventional and cellular automata-type games, where in the latter simulating evolution agents 'die', and are occasionally replaced; in our simulations we used only 'conventional' games, where agent-replacement was not allowed).

**Table S1.2.** Effect of learning on cooperation

| Type of learning[a] | Effect on cooperation | Networks; games; strategy adoption rules | Agents (players) | References |
|---|---|---|---|---|
| **One-step learning strategy adoption rules** | **Help** (increases cooperation in repeated multi-agent games) | Lattice; PD[b]; Tit-for-tat strategy adoption rule and its generous versions[c] | Simulation | 61–63 |
| **Two-step learning strategy adoption rules**[d] | **Help** (make cooperation rather resistant to noise → often win against Tit-for-tat) | Lattice; PD; Pavlov strategy adoption rule and its generous versions[c] | Simulation | 61–66 |
| **Extended learning strategy adoption rules** (3 or more steps) | **Help** (each additional memory unit contributes less to the increase of cooperation) | Lattice, scale-free; HD[b], PD, alternating PD with noise; higher memory 'Firm Pavlov', 'Meta-Pavlov' strategy adoption rules | Simulation | 32, 62, 67–71 |
| **Complex learning strategy adoption rules** (adaptive learning, operant conditioning, preferential learning, Q-learning, reinforcement learning) | **Help** (are not only resistant to noise but can exploit noise to drift towards cooperation, reinforcement learning based on local or global information enables sophisticated strategy adoption rules to emerge and allows efficient network formation) | Lattice, scale-free; HD, matching pennies game, PD; pair-wise comparison dynamics strategy adoption rule | Simulation, primates, humans | 12, 36, 37, 46, 72–77 |
| **Natural learning processes** | **Help** (fishes, monkeys remember their cooperators; birds learn cooperation with feedback signals or accumulated payoffs; lions learn cooperative hunting to capture fast prey; vampire bats share blood by regurgitation; students are more successful using complex Pavlov strategy adoption rules than tit-for-tat, which is the default, if their memory capacity is compromised disfavoring cooperation; subjects with psychopathy disorders have a deficit of emotional reward for cooperation, which can be corrected by learning) | PD (interfering Memory game) | Guppies, birds, vampire bats, lions, monkeys, humans (controls, subjects with psychopathy, autism, or attention-deficit hyperactivity disorder) | 33, 78–86 |

**Table S1.2.** Effect of learning on cooperation (continued)

| Type of learning[a] | Effect on cooperation | Networks; games; strategy adoption rules | Agents (players) | References |
|---|---|---|---|---|
| **Communication, negotiation** | **Help** (viruses lack communication and cooperation; quorum sensing is required for bacterial biofilm formation; avoidance of discussion blocks cooperation; complex communication allows better cooperation; feedback eases internet and traffic congestion; firm's market image helps cooperative response; description of future goals greatly enhances cooperation) | PD, 'game of sexes', biofilm formation, internet usage, car-race, trade | Simulation, viruses, bacteria, humans, firms | 41, 72, 73, 87–94 |
| **Quantum entanglement** ('quantum communication') | **Helps** (quantum bits, 'qubits' enable a continuous cooperation, which works as a contract) | Quantum minority game, quantum PD | Simulation | 95 |
| **Tag, reputation-building** | **Help** (establishing and learning tags and reputation help cooperators to detect each other – even without memory – and build communities) | Donation game, PD, ultimatum game, car-race, e-trade | Simulation, humans | 19, 27, 41, 67, 96–98 |
| **Evolutionary preserved recognition** (using the Baldwin-effect) | **Helps** (enables the detection and avoidance of cheaters; the learned habit is selected and fixed by evolution) | Hermaphrodites exchanging eggs | Hermaphrodite worms | 99 |
| **Memory of cooperation patterns** (cultural context) | **Helps** (cooperation in previous games; cooperative educational or cultural traits) | Intergenerational public good game, PD, ultimatum game | Humans | 12, 27, 100, 101 |

[a] The term 'learning' is used here in the sense of the collection and use of information influencing game strategy adoption rules and behavior, and not in the restricted sense of imitation, or directed information-flow from a dominant source (the teacher). Therefore, learning here includes communication, negotiation, memory, label-assignment and label-recognition, etc.

[b] HD = Hawk-Dove (Snowdrift, Chicken) game; PD = Prisoner's Dilemma game (please note that in this supplementary table we did not discriminate between conventional and cellular automata-type games, where in the latter simulating evolution agents 'die', and are occasionally replaced; in our simulations we used only 'conventional' games, where agent-replacement was not allowed).

[c] Tit-for-tat = this strategy adoption rule copies the opponent's step in the previous round; Pavlov = a 'win stay – lose shift' strategy adoption rule; generous strategy adoption rules = allow 'extra' cooperation options with a given probability.

[d] These strategy adoption rules are interchangeably called as 'memory-one' or 'memory-two' strategy adoption rules referring to the fact that e.g. in the Pavlov strategy adoption rule agents remember the outcome of only the last step ('memory-one') but that of both players ('memory-two').

**Table S1.3.** Effect of innovation on cooperation

| Type of innovation[a] | Effect on cooperation | Networks; games; strategy adoption rules | Agents (players) | References |
|---|---|---|---|---|
| **Topological irregularities** (empty sites = 'sterile defectors', small-world shortcuts, hubs) | **Mostly help** (see Table 1, block the spread of defection, however high degree *inhibits* cooperation and irregularities make it *sensitive* for strategy adoption rules) | Lattice, small-world; HD, PD[b] | Simulation | 49, 102 and Table S1.1 |
| **Low noise** (random noise, errors, mistakes, the 'trembling hand') | **Helps** (at low levels resolves deadlocks, at high levels *inhibits* cooperation) | Evolutionary language learning game, ultimatum game | Simulation | 42, 98 |
| **High noise** (random noise, errors, mistakes, the 'trembling hand') | **Inhibits** (PD game is noise-sensitive, especially on lattices, where noise makes cooperator boundaries irregular) | Lattice; PD | Simulation | 52, 102–104 |
| **Pink noise** (chaotic changes in environment affecting payoff) | **Mostly helps** (smaller, but reliable payoffs become more attractive) | Lattice; PD | Simulation | 105 |
| **Random elements in strategy adoption rules** (strategy selection, payoff determination, etc.) | **Help** (at low levels resolve deadlocks, at high levels inhibit cooperation) | Lattice, random, small-world; HD, PD | Simulation | 48, 50, 55, 106, 107 |
| **Random extra cooperation in strategy adoption rules** | **Helps** | Lattice; PD; Generous tit-for-tat, 'double-generous-tit-for-tat' | Simulation | 65, 108 |
| **Mutation of strategy adoption rules** | **Helps** (may re-introduce cooperation) | Lattice; PD | Simulation | 66 |
| **Extra loner strategy adoption rule**[c] | **Helps** (even for large temptation values) | Lattice, small-world; PD, public good game | Simulation | 107, 109–112 |
| **Quantum probabilistic strategies** | **Help** (ancillary quantum bits, 'qubits' enable to use 'mixed' strategies) | Quantum minority game, quantum PD | Simulation | 95 |
| **Random elements in strategy adoption rules** | **Help** (increased when playing games) | matching pennies game, PD and other social dilemma games | Simulation, humans, primates | 34–37, 113 |
| **Mixed strategies** | **Help** (reputation building is supplemented with costly punishment) | PD | Humans | 114 |
| **Egalitarian motives** | **Help** (help the development of reciprocity) | Public good game | Humans | 40 |

[a]The term 'innovation' is used here in the sense of irregularities in the process of the game. Therefore, innovation here includes errors, mutations, mistakes, noise, randomness and increased temperature besides the *senso stricto* innovation of conscious, intelligent agents.

[b]HD = Hawk-Dove (Snowdrift, Chicken) game; PD = Prisoner's Dilemma game (please note that in this supplementary table we did not discriminate between conventional and cellular automata-type games, where in the latter simulating evolution agents 'die', and are occasionally replaced; in our simulations we used only 'conventional' games, where agent-replacement was not allowed).

[c]Loners do not participate in the game and share the income with the co-player.

## Supplementary Figures

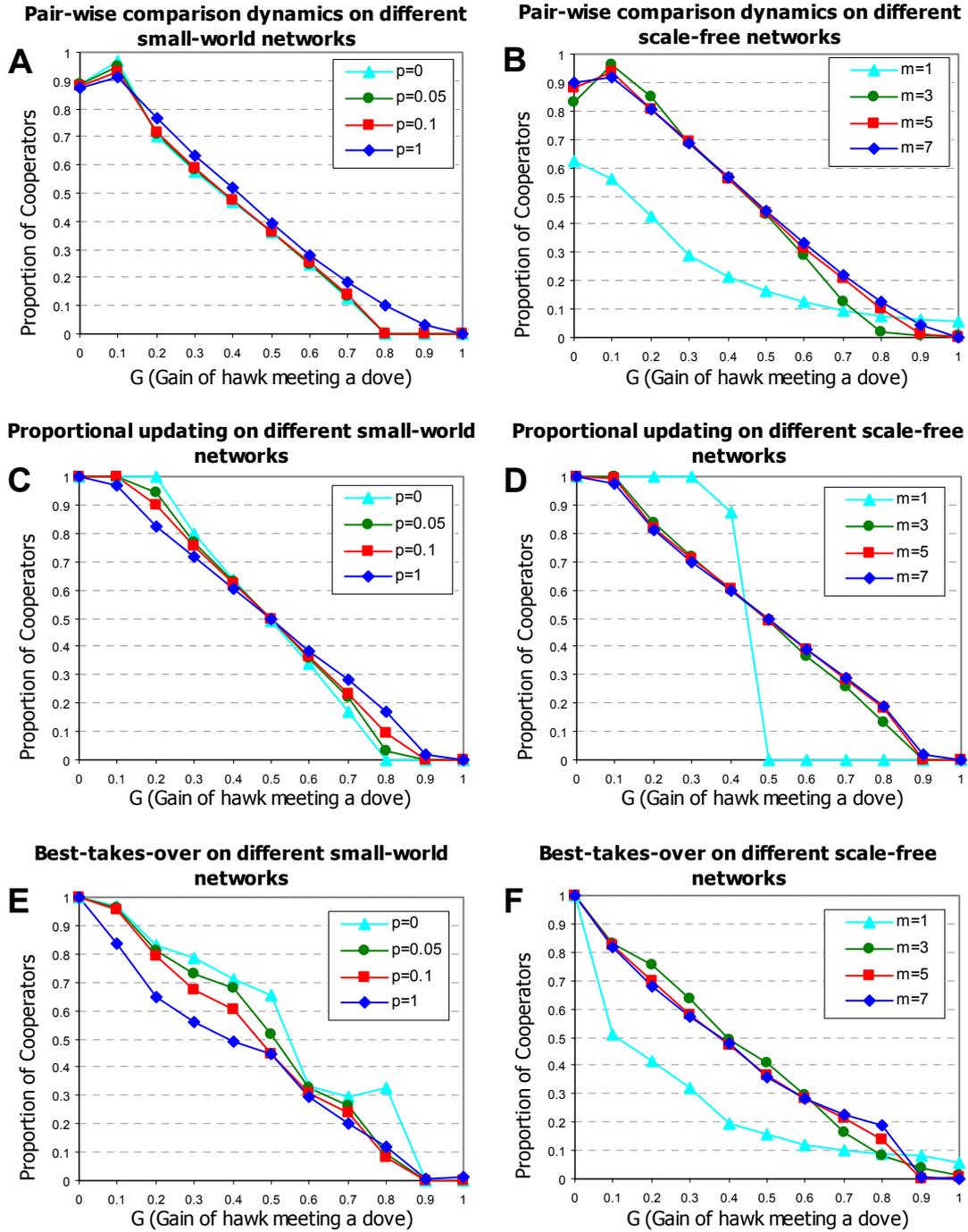

**Figure S1.1.** Variation of cooperation level using short-term, non-innovative strategy adoption rules in Hawk-Dove games on small-world and scale-free networks. The modified Watts-Strogatz small-world networks (Panels *A, C* and *E*) were built on a 50 x 50 lattice, where each node was connected to its eight nearest neighbors. The rewiring probability of the regular links was 0 (pale blue triangles), 0.05 (green circles), 0.1 (red squares) and 1 (dark blue diamonds). The Barabasi-Albert scale-free networks (Panels *B, D* and *F*) also contained 2,500 nodes, where at each construction step a new node was added with m=1 (pale blue triangles), m=3 (green circles), m=5 (red squares) or m=7 (dark blue diamonds) new links attached to the existing nodes. For the description of the networks, Hawk-Dove games and the three different strategy adoption rules, the pair-wise comparison dynamics (Panels *A* and *B*), the proportional updating (Panels *C* and *D*) and the best-takes-over strategy adoption rules (Panels *E* and *F*), see Methods. For each strategy adoption rule and *G* values (representing the gain of hawk meeting a dove, see Methods), 100 random runs of 5,000 time steps were executed.

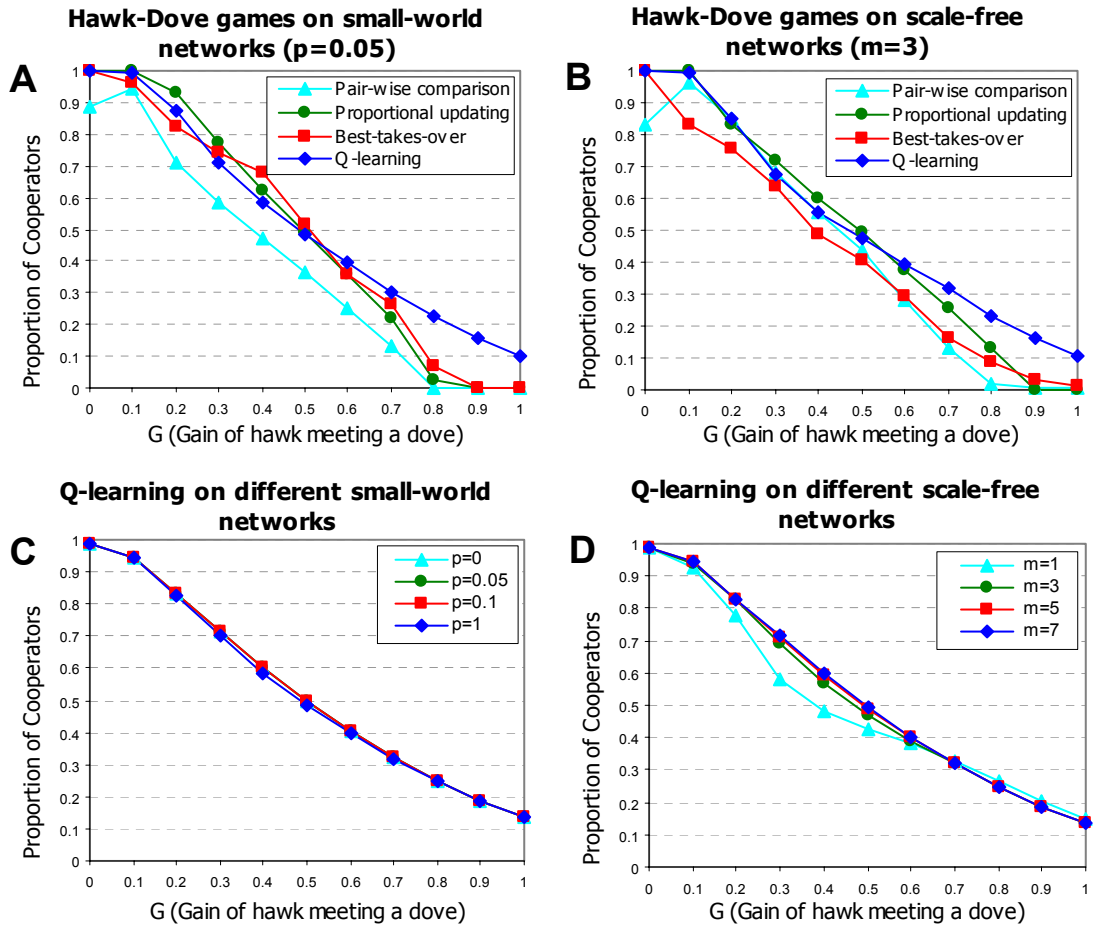

**Figure S1.2.** Q-learning improves and stabilizes the cooperation of agents forming small-world and scale-free networks in Hawk-Dove games. *A*, The modified Watts-Strogatz small-world networks [2] were built on a 50 x 50 lattice, where each node was connected to its eight nearest neighbors. The rewiring probability of the regular links was 0.05. *B*, The Barabasi-Albert scale-free networks [10] also contained 2,500 nodes, where at each construction step a new node was added with m=3 new links attached to the existing nodes. For the description of the Hawk-Dove games and the four different strategy adoption rules, pair-wise comparison dynamics (pale blue triangles), proportional updating (green circles), best-takes-over (red squares) and Q-learning (dark blue diamonds) see Methods. *C*, The rewiring probability of the small-world network of panel *A* was 0 (regular network, pale blue triangles), 0.05 (small-world, green circles), 0.1 (small-world, red rectangles) and 1 (random network, dark blue diamonds). *D*, The number of nodes added to the existing nodes of the scale-free network of *B* was varied between 1 and 7. For each strategy adoption rule and *G* values (representing the gain of hawk meeting a dove, see Methods), 100 random runs of 5,000 time steps were executed.

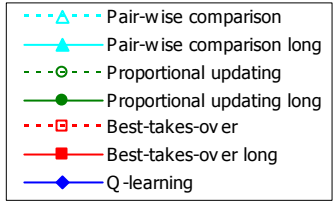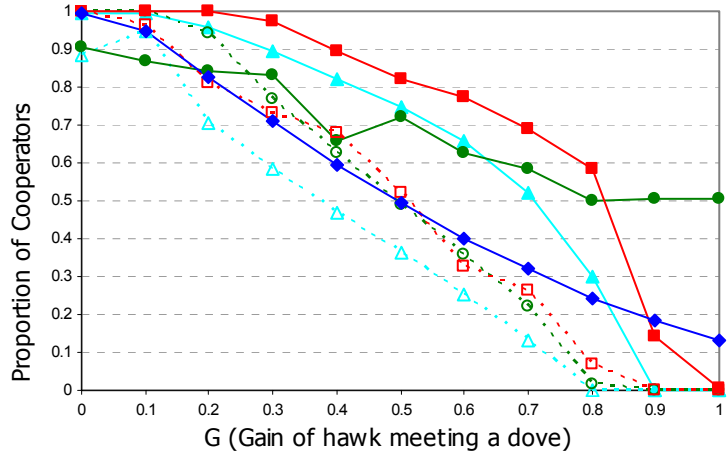

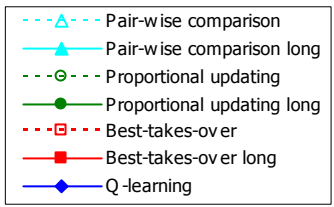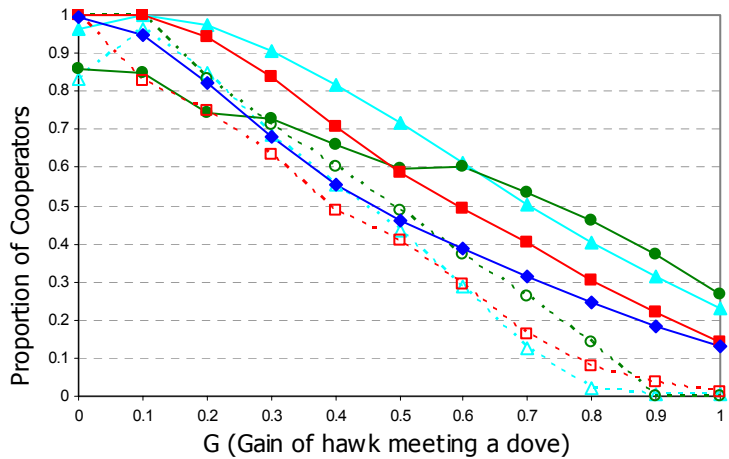

**Figure S1.3.** Long-term learning strategy adoption rules help cooperation in Hawk-Dove games played on various networks. For the description of the small-world [2] and scale-free [10] networks, the Hawk-Dove game and the different strategy adoption rules, pair-wise comparison dynamics (pale blue open triangles and dashed line), proportional updating (green open circles and dashed line), best-takes-over (red open squares and dashed line), Q-learning (dark blue diamonds and solid line) pair-wise comparison dynamics long (pale blue filled triangles and solid line), proportional updating long (green filled circles and solid line) and best-takes-over long (red filled squares and solid line) strategy adoption rules see Methods. *A*, Long-term learning strategy adoption rules on small-world networks with a rewiring probability of 0.05. *B*, Long-term learning strategy adoption rules on scale-free networks with m=3. For each game strategy adoption rule and *G* values (representing the gain of hawk meeting a dove, see Methods), 100 random runs of 5,000 time steps were executed.

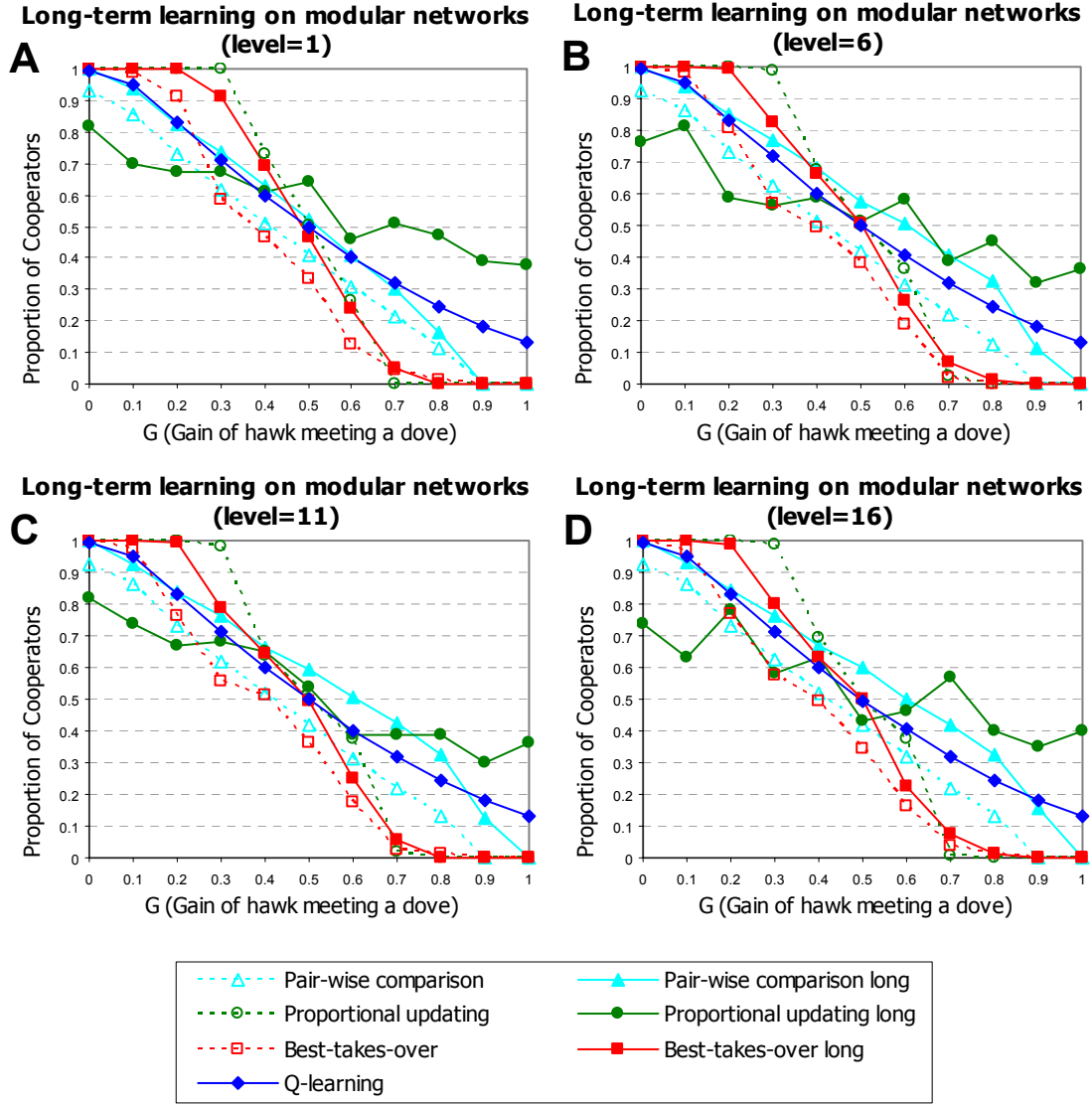

**Figure S1.4.** Long-term learning strategy adoption rules help cooperation in Hawk-Dove games played on modular networks. In the modular networks described by Girvan and Newman [11] each network had a scale-free degree distribution, contained 128 nodes and was divided into 4 communities. The average degree was 16. Panels *A* through *D* show the % of cooperation when playing on Girvan-Newman modular networks with levels 1, 5, 10 or 16, respectively, where 'level 1' means that for each node in the network, the expected number of links between a node and the nodes which are in other communities was 1. With increasing 'level' the community structure died down gradually. For the description of the Hawk-Dove game and the different strategy adoption rules**,** pair-wise comparison dynamics (pale blue open triangles and dashed line), proportional updating (green open circles and dashed line), best-takes-over (red open squares and dashed line), Q-learning (dark blue filled diamonds and solid line) pair-wise comparison dynamics long (pale blue filled triangles and solid line), proportional updating long (green filled circles and solid line) and best-takes-over long (red filled squares and solid line) strategy adoption rules see Methods. For each game strategy adoption rule and *G* values runs on 100 Girvan-Newman-type modular networks of 5,000 time steps were executed.

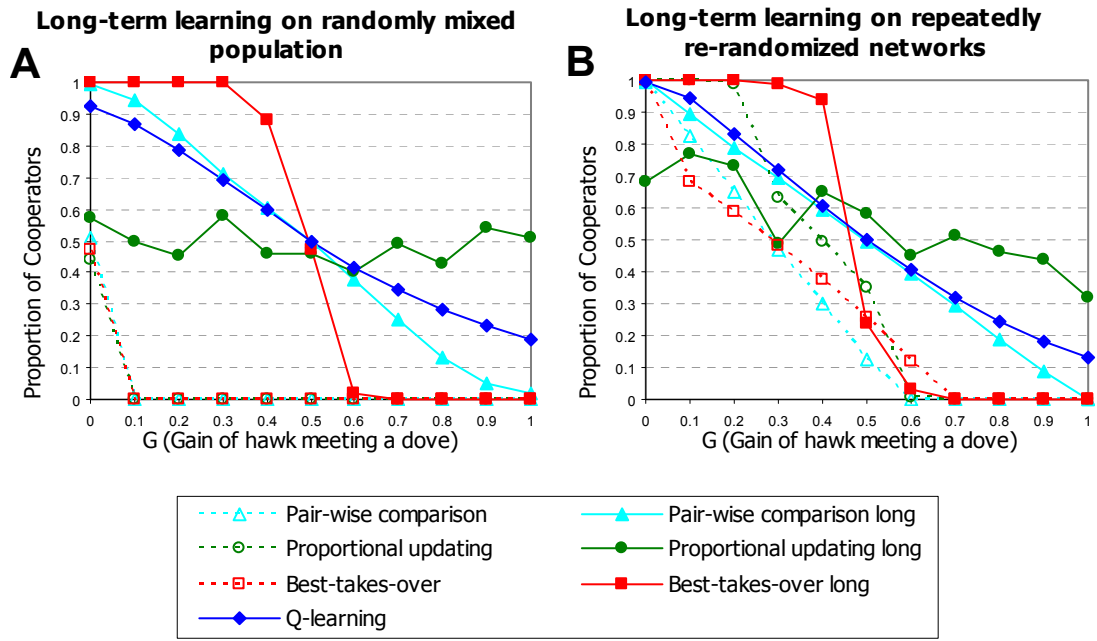

**Figure S1.5.** Long-term learning strategy adoption rules help cooperation in Hawk-Dove games on randomly mixed population and on repeatedly re-randomized networks. For the description of the Hawk-Dove game and the different strategy adoption rules, pair-wise comparison dynamics (pale blue open triangles and dashed line), proportional updating (green open circles and dashed line), best-takes-over (red open squares and dashed line), Q-learning (dark blue filled diamonds and solid line) pair-wise comparison dynamics long (pale blue filled triangles and solid line), proportional updating long (green filled circles and solid line) and best-takes-over long (red filled squares and solid line) strategy adoption rules see Methods. *A*, Games between two randomly selected agents from 100 total. For each game strategy adoption rule and *G* values, 100 random runs of 100,000 time steps were executed. *B*, Before each individual rounds of the repeated Hawk-Dove game, we generated a new random graph of the agents with a connection probability, p=0.02, where the number of agents was 200. In this way for a specific agent, its neighbors changed in each round of game. For each game strategy adoption rule and *G* values (representing the gain of hawk meeting a dove, see Methods), 100 random runs of 5,000 time steps were executed.

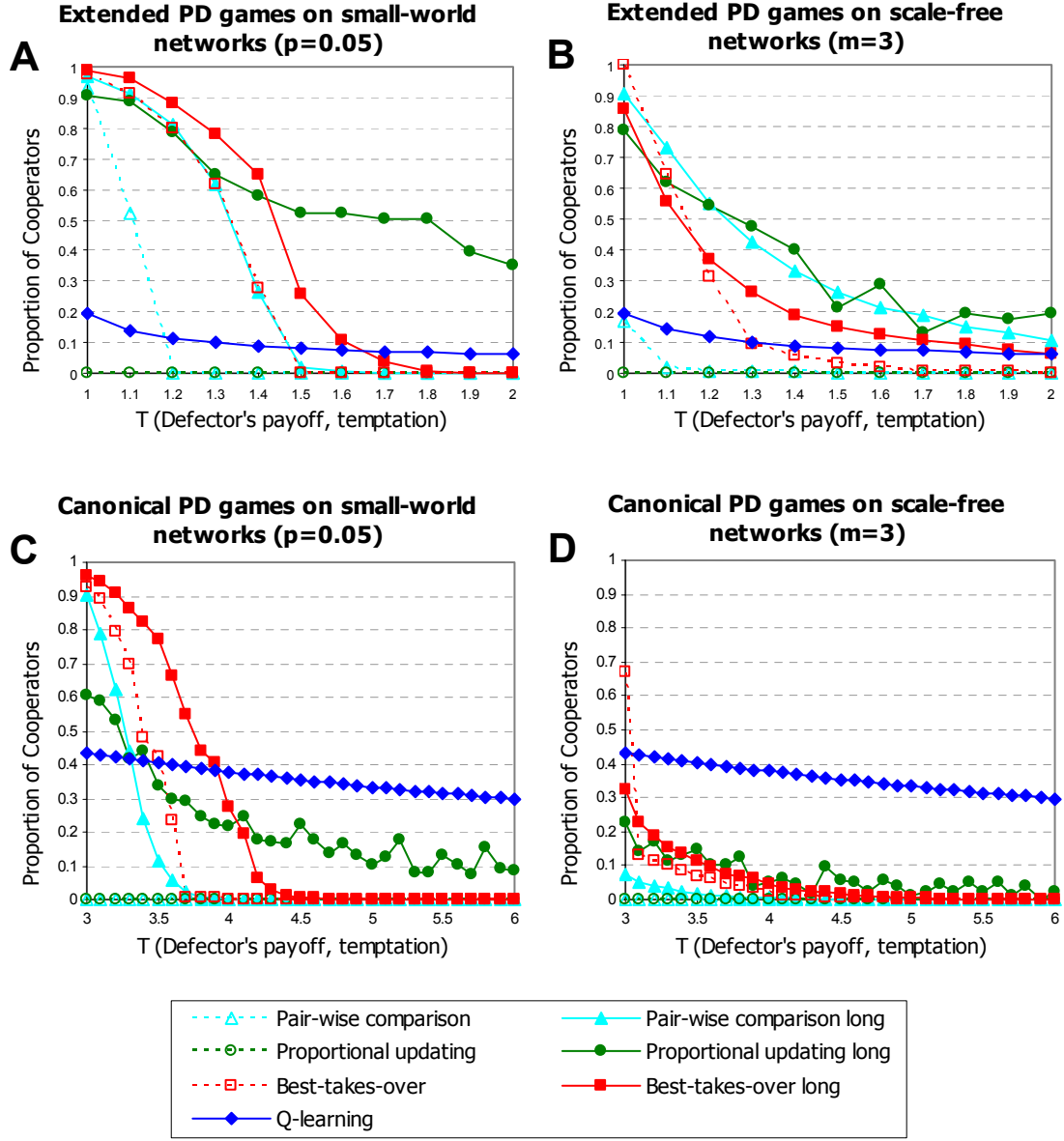

**Figure S1.6.** Long-term learning strategy adoption rules help cooperation in both canonical and extended Prisoner's Dilemma games played on small-world and scale-free networks. The small-world (panels *A* and *C*, [2]) and scale-free (panels *B* and *D*, [10]) networks were built as described in the Methods. For the description of the Prisoner's Dilemma games and the different strategy adoption rules, pair-wise comparison dynamics (pale blue open triangles and dashed line), proportional updating (green open circles and dashed line), best-takes-over (red open squares and dashed line), Q-learning (dark blue filled diamonds and solid line) pair-wise comparison dynamics long (pale blue filled triangles and solid line), proportional updating long (green filled circles and solid line) and best-takes-over long (red filled squares and solid line) strategy adoption rules see Methods. Panels *A* and *B*, extended Prisoner's Dilemma games ( $R=1, P=0, S=0$ *T* was changed from 1 to 2; 1). Panels *C* and *D*, canonical Prisoner's Dilemma games ( $R=3, P=1, S=0$ *T* was changed from 3 to 6; [6]). In the canonical Prisoner's Dilemma games when using the Q-learning, the initial annealing temperature was set to 10,000 to extend the annealing process [115]). For each game strategy adoption rule and *T* values 100 random runs of 5,000 time steps were executed.

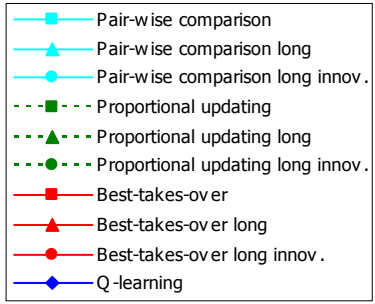
**A** Hawk-Dove games on small-world networks (p=0.05)

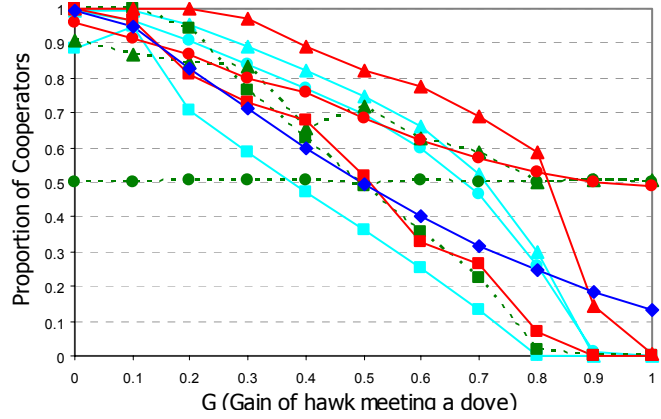

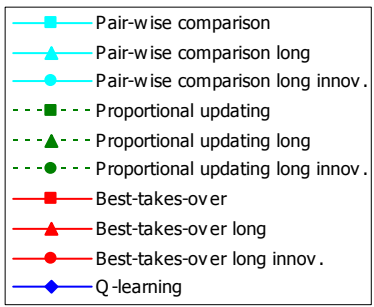
**B** Hawk-Dove games on scale-free networks (m=3)

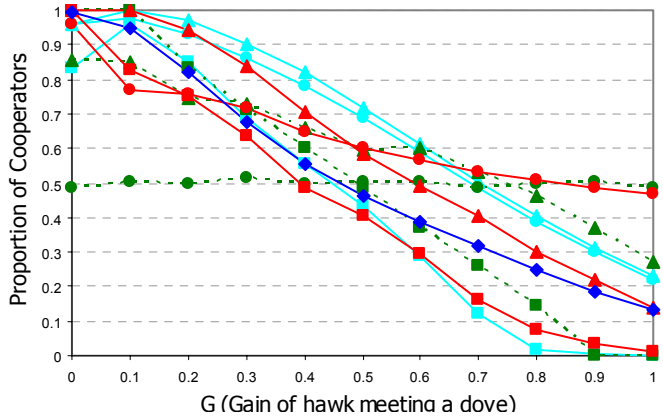

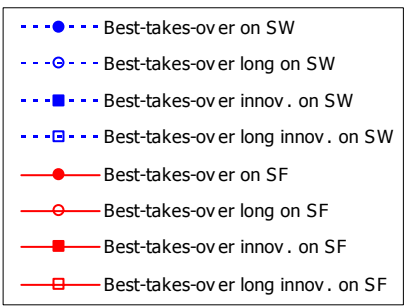
**C** Best-takes-over on small-world (p=0.05) and scale-free (m=3) networks

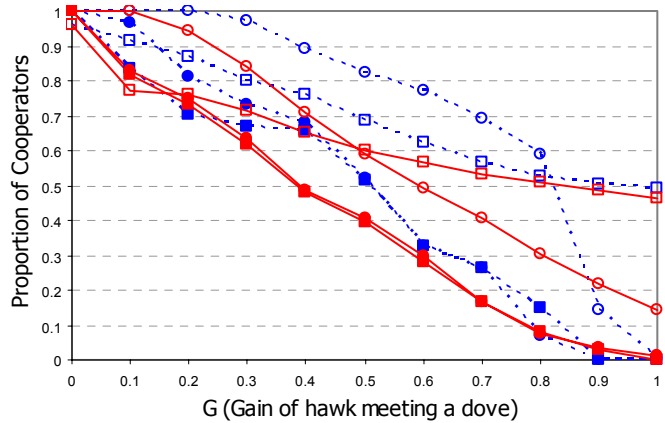

**Figure S1.7.** Comparison of innovative strategy adoption rules in Hawk-Dove games on small-world and scale-free networks. The small-world (*A* and *C* blue symbols and dashed lines) and scale-free (*B* and *C* red symbols and solid lines) networks were built as described in Methods. For the description of the Hawk-Dove game and the different strategy adoption rules, pair-wise comparison dynamics (pale blue filled squares, solid line), pair-wise comparison dynamics long (pale blue filled triangles, solid line), pair-wise comparison dynamics long innovative (pale blue filled circles, solid line), proportional updating (green filled squares, dashed line), proportional updating long (green filled triangles, dashed line), proportional updating long innovative (green filled circles, dashed line), best-takes-over (on panel *A* and *B*: red filled squares, on panel *C*: filled circles), best-takes-over long (on panel *A* and *B*: red filled triangles, on panel *C*: open circles), best-takes-over innovative (on panel *C*: filled squares), best-takes-over long innovative (on panel *A* and *B*: red filled circles, on panel *C*: open squares), and Q-learning (blue filled diamonds) strategy adoption rules, see Methods. For each game strategy adoption rule and *G* values (representing the gain of hawk meeting a dove, see Methods), 100 random runs of 5,000 time steps were executed.

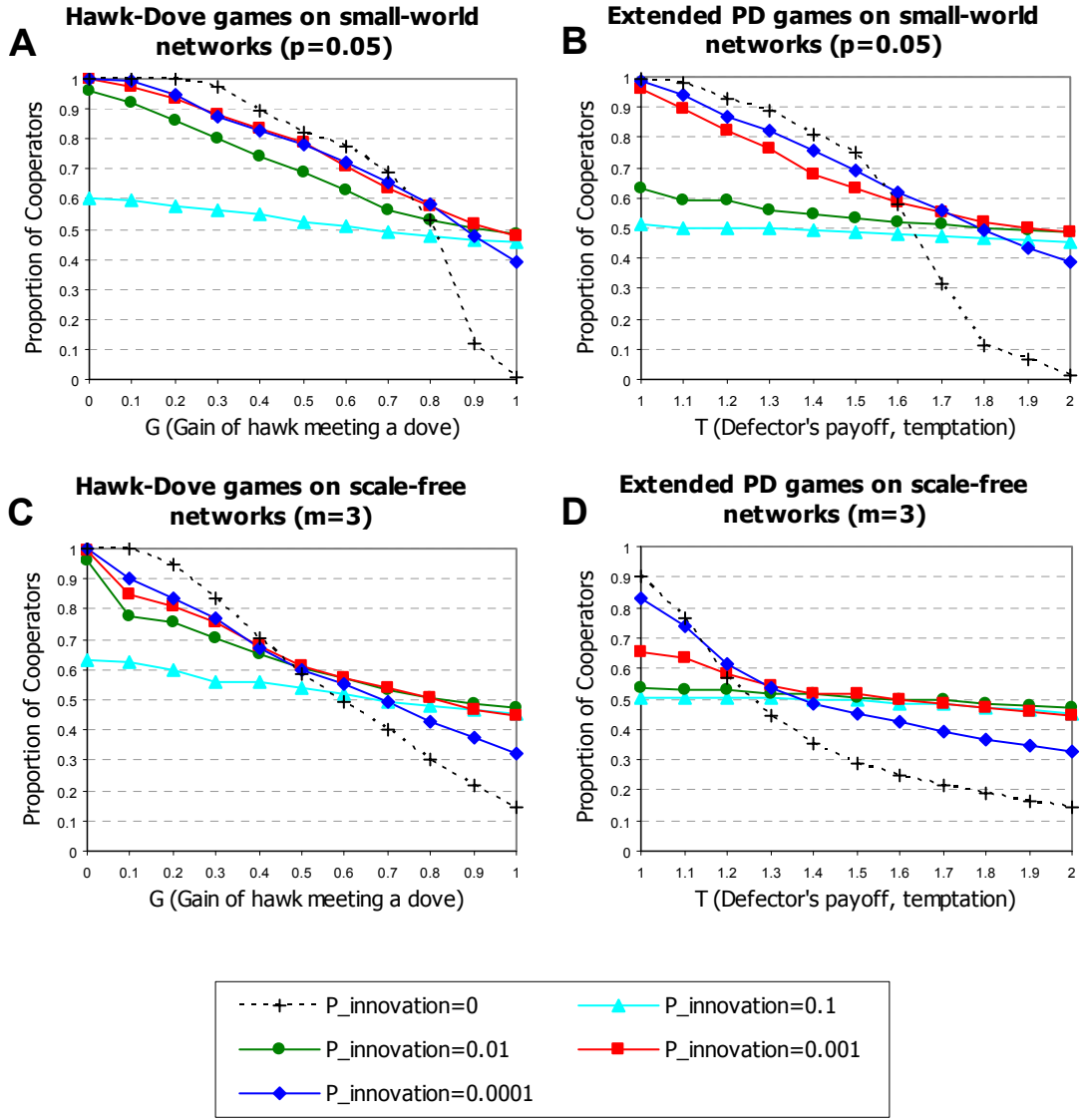

**Figure S1.8.** Comparison of different innovation levels of the best-takes-over long innovative strategy adoption rule in Hawk-Dove and extended Prisoner's Dilemma games on small-world and scale-free networks. The small-world (panels *A* and *B*, [2]) and scale-free (panels *C* and *D*, [10]) networks were built as described in the Methods. For the description of the Hawk-Dove game (panels *A* and *C*), extended Prisoner's Dilemma game (panels *B* and *D*) and the best-takes-over long innovative strategy adoption rule, see Methods. The probability of innovation was changed from zero to 0.1 as described in the Figure legend. For each game strategy adoption rule and *G* values (representing the gain of hawk meeting a dove, see Methods), 100 random runs of 5,000 time steps were executed.

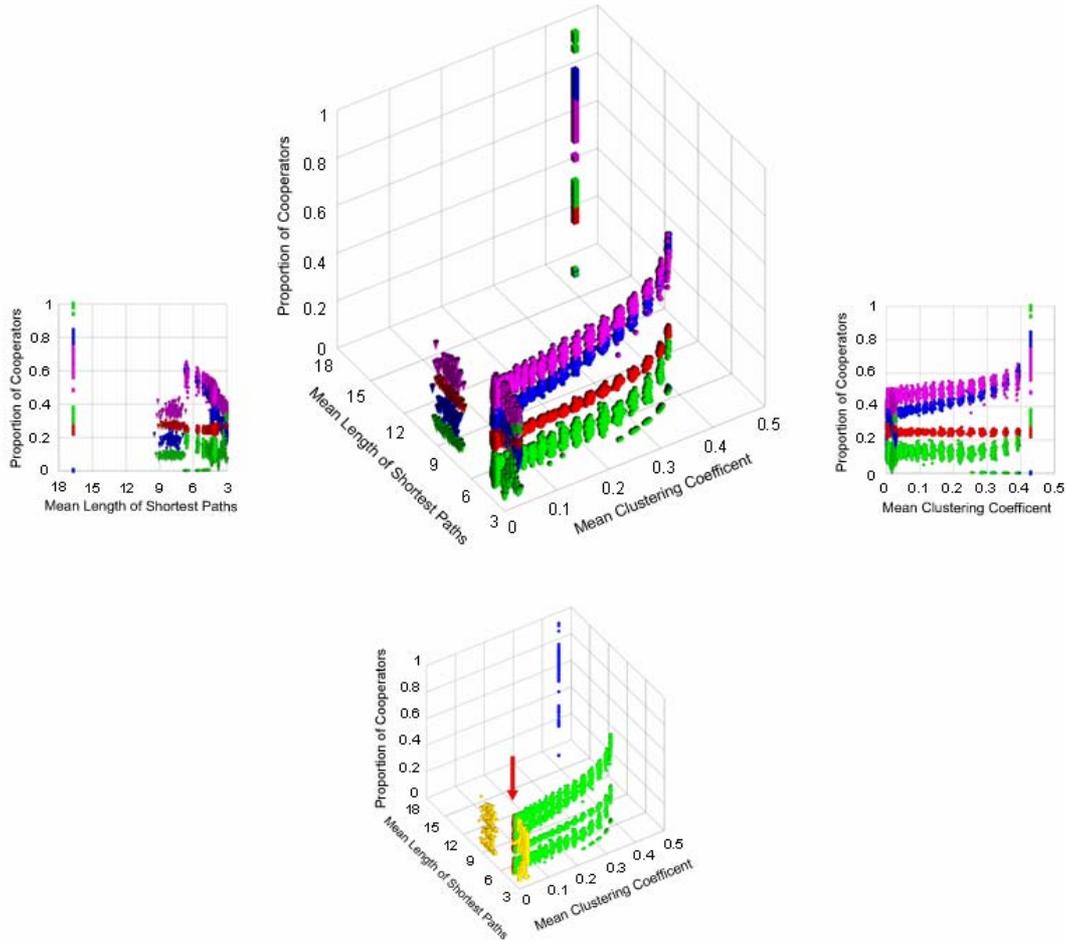

**Figure S1.9.** Long-term learning and innovative strategy adoption rules extend cooperative network topologies in the Hawk-Dove game. The *top middle panel* shows the level of cooperation at different network topologies. small-world (spheres) and scale-free (cones) networks were built as described in the Methods. The rewiring probability, *p* of small-world networks was increased from 0 to 1 with 0.05 increments, the number of edges linking each new node to former nodes in scale-free networks was varied from 1 to 7, and the means of shortest path-lengths and clustering coefficients were calculated for each network. Cubes and cylinders denote regular (p = 0) and random (p = 1.0) extremes of the small-world networks, respectively. For the description of the games and the best-takes-over (green symbols); long-term learning best-takes-over (blue symbols); long-term learning innovative best-takes-over (magenta symbols) and Q-learning (red symbols) strategy adoption rules used, see Methods. The *left and right panels* show the 2D side views of the 3D top middle panel using the same symbol-set. For each network 100 random runs of 5,000 time steps were executed at a fixed *G* value of 0.8. The *bottom middle panel* shows a color-coded illustration of the various network topologies used on the top middle panel. Here the same simulations are shown as on the top middle panel with a different color-code emphasizing the different network topologies. The various networks are represented by the following colors: regular networks – blue; small-world networks – green; scale-free networks – yellow; random networks – red (from the angle of the figure the random networks are behind some of the small-world networks and, therefore are highlighted with a red arrow to make there identification easier).

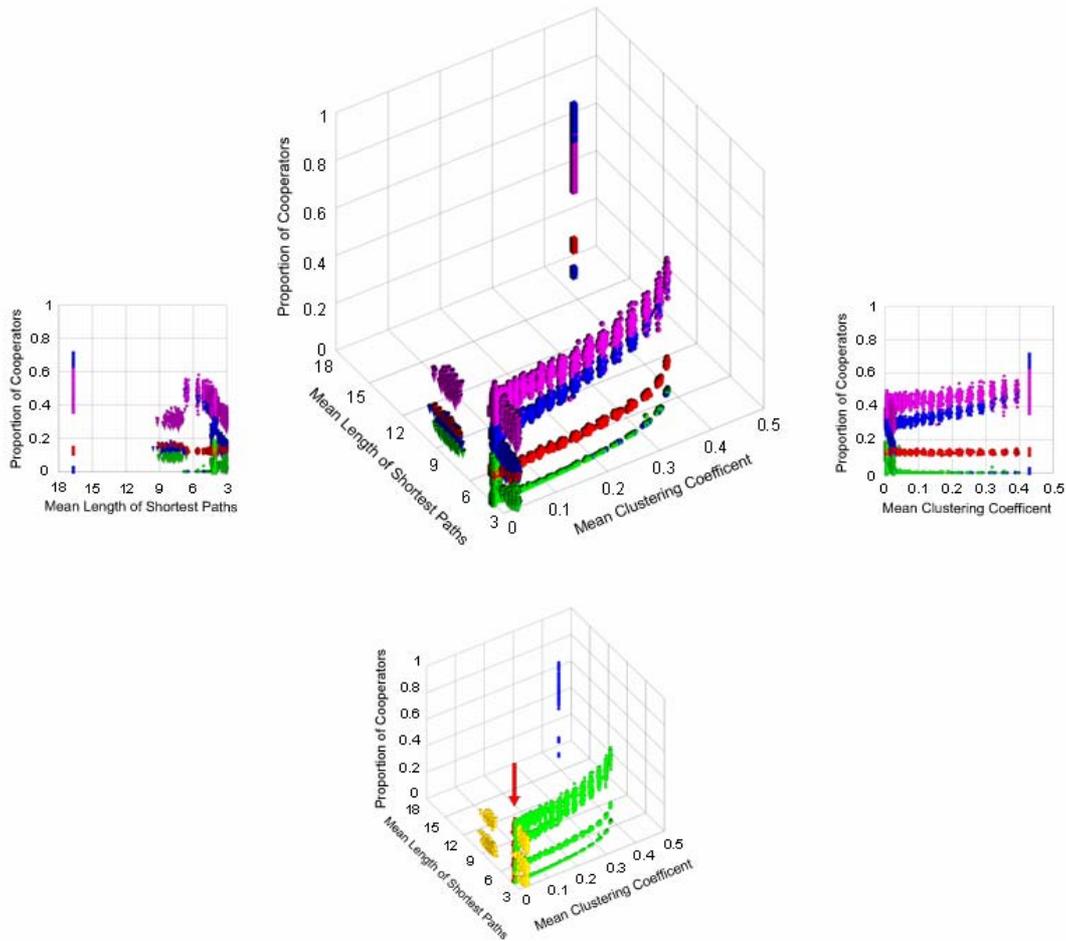

**Figure S1.10.** Long-term learning and innovative strategy adoption rules extend cooperative network topologies in the extended Prisoner's Dilemma game. The *top middle panel* shows the level of cooperation at different network topologies. small-world (spheres) and scale-free (cones) networks were built as described in the Methods. The rewiring probability, *p* of small-world networks was increased from 0 to 1 with 0.05 increments, the number of edges linking each new node to former nodes in scale-free networks was varied from 1 to 7, and the means of shortest path-lengths and clustering coefficients were calculated for each network. Cubes and cylinders denote regular (p = 0) and random (p = 1.0) extremes of the small-world networks, respectively. For the description of the games and the best-takes-over (green symbols); long-term learning best-takes-over (blue symbols); long-term learning innovative best-takes-over (magenta symbols) and Q-learning (red symbols) strategy adoption rules used, see Methods. The *left and right panels* show the 2D side views of the 3D top middle panel using the same symbol-set. For each network 100 random runs of 5,000 time steps were executed at a fixed *T* value of 1.8. The *bottom middle panel* shows a color-coded illustration of the various network topologies used on the top middle panel. Here the same simulations are shown as on the top middle panel with a different color-code emphasizing the different network topologies. The various networks are represented by the following colors: regular networks – blue; small-world networks – green; scale-free networks – yellow; random networks – red (from the angle of the figure the random networks are behind some of the small-world networks and, therefore are highlighted with a red arrow to make there identification easier).

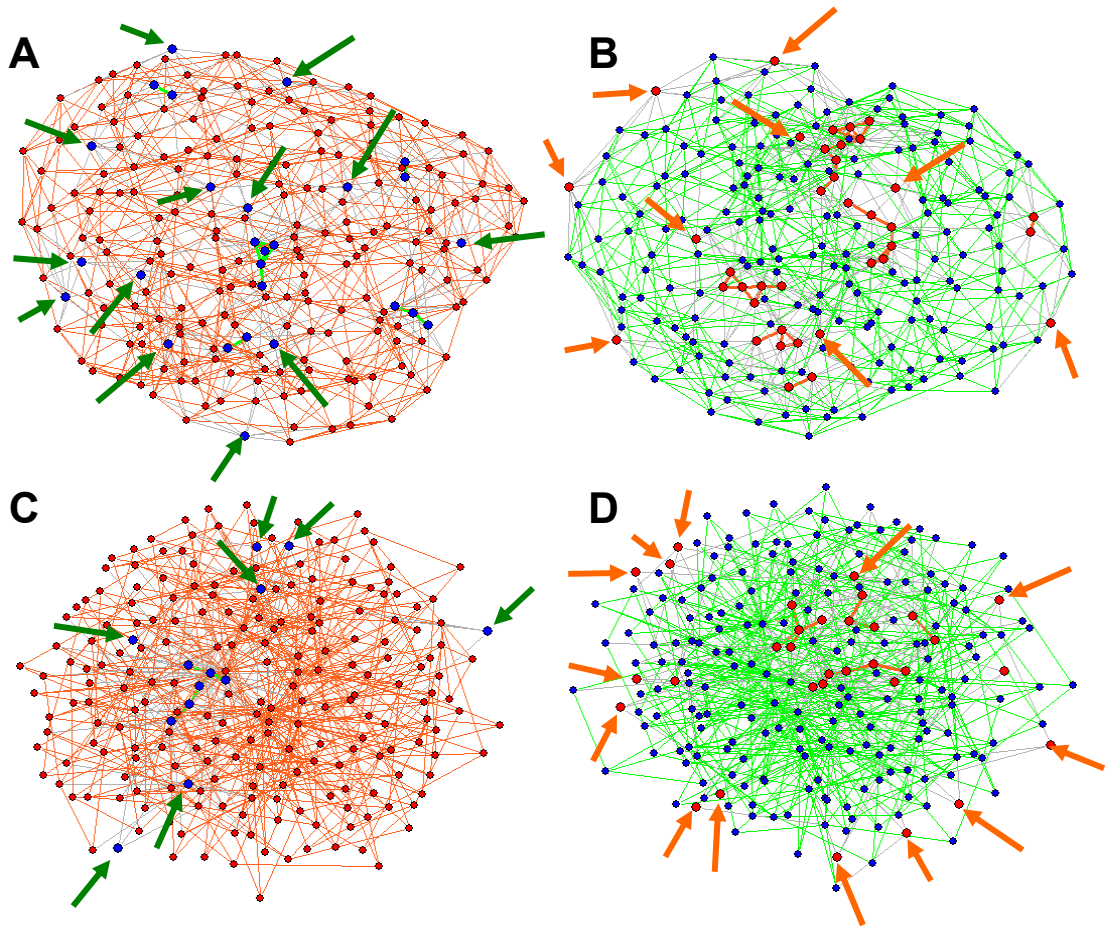

**Figure S1.11.** Both hawks and doves become isolated in extreme minority, when they use the innovative Q-learning strategy adoption rule in Hawk-Dove games on small-world and scale-free networks. The small-world [2] and scale-free networks [10] were built, and Hawk-Dove games were played as described in the Methods using 225 agents. Networks showing the last round of 5,000 plays were visualized using the Kamada-Kawai algorithm of the Pajek program [116]. Blue and orange dots correspond to hawks and doves, respectively. Green, orange and grey lines denote hawk-hawk, dove-dove or dove-hawk contacts, respectively. Arrows point to lonely hawks or doves using the respective colors above. *A*, Small-world network with a rewiring probability of 0.05, *G*=0.15. *B*, Small-world network with a rewiring probability of 0.05, *G*=0.95. *C*, Scale-free network with m=3, *G*=0.1. *D*, Scale-free network with m=3, *G*=0.98. We have received similar data when playing extended Prisoner's Dilemma games (data not shown).

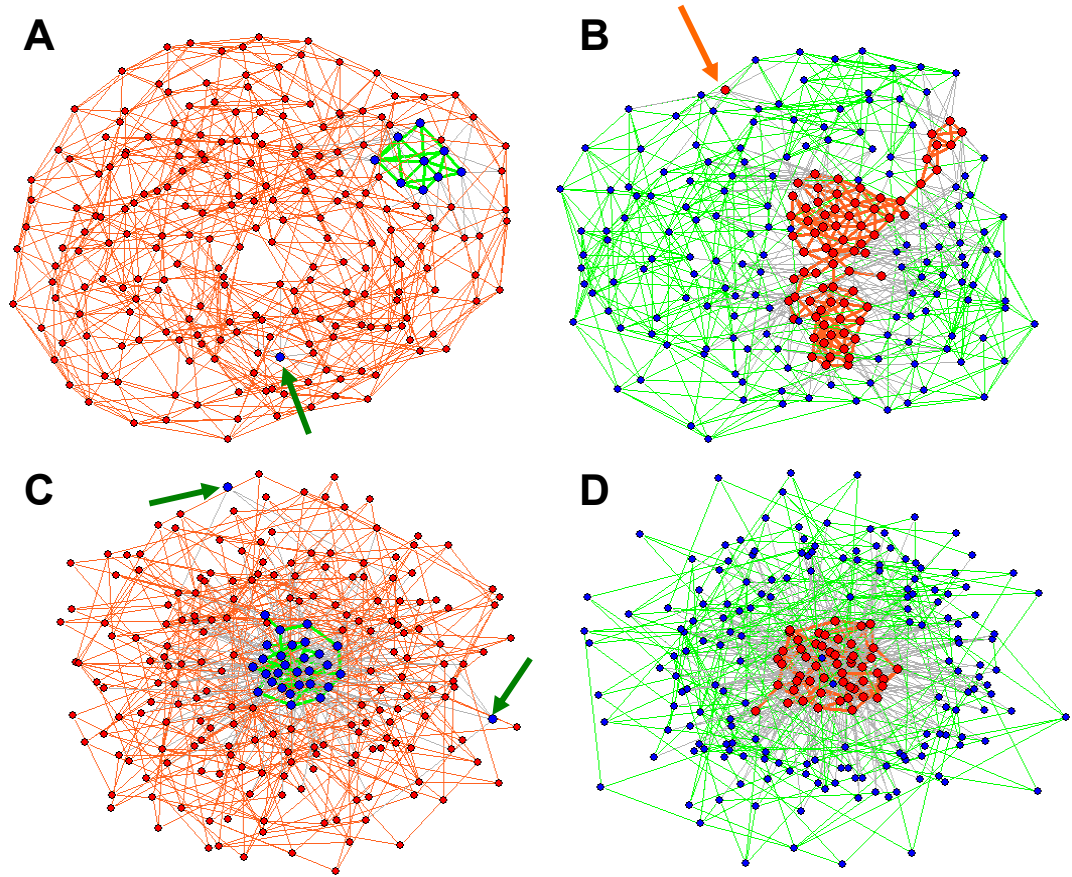

**Figure S1.12.** Hawks, and especially doves are not extremely isolated in extreme minority, when they use the non-innovative best-takes-over strategy adoption rule in Hawk-Dove games on small-world and scale-free networks. The small-world [2] and scale-free networks [10] were built, and Hawk-Dove games were played as described in the Methods using 225 agents. Networks showing the last round of 5,000 plays were visualized using the Kamada-Kawai algorithm of the Pajek program [116]. Blue and orange dots correspond to hawks and doves, respectively. Green, orange and grey lines denote hawk-hawk, dove-dove or dove-hawk contacts, respectively. Arrows point to lonely hawks or doves using the respective colours above. *A*, Small-world network with a rewiring probability of 0.05, *G*=0.15. *B*, Small-world network with a rewiring probability of 0.05, *G*=0.75. *C*, Scale-free network with m=3, *G*=0.1. *D*, Scale-free network with m=3, *G*=0.8. We have received similar data using other non-innovative strategy adoption rules, such as pair-wise comparison dynamics, or proportional updating, as well as when playing extended Prisoner's Dilemma games (data not shown).